\documentclass[lettersize,journal]{IEEEtran}
\usepackage{amsmath,amsfonts}
\usepackage{algorithmic}
\usepackage{algorithm}
\usepackage{array}
\usepackage[caption=false,font=footnotesize]{subfig}
\usepackage{textcomp}
\usepackage{hyperref}
\usepackage{tabularx}
\usepackage{stfloats}
\usepackage{url}
\usepackage{verbatim}
\usepackage{graphicx}
\usepackage{cite}
\usepackage{multirow}
\usepackage{color}
\begin{document}

\title{Multi-Modal Sensing Residual-Corrected GNN for mmWave Path Loss Prediction via Synesthesia of Machines}

\author{Mengyuan Lu,~\IEEEmembership{Member,~IEEE}, Lu~Bai,~\IEEEmembership{Senior Member,~IEEE}, and Xiang~Cheng,~\IEEEmembership{Fellow,~IEEE}
        % <-this % stops a space
\thanks{M.~Lu is with the Joint SDU-NTU Centre for Artificial Intelligence Research (C-FAIR), Shandong University, Jinan, 250101, P. R. China, and also with the School of Software, Shandong University, Jinan 250101, P. R. China (e-mail: mengyuanlu@mail.sdu.edu.cn).}
\thanks{L.~Bai is with the Joint SDU-NTU Centre for Artificial Intelligence Research (C-FAIR), Shandong University, Jinan, 250101, P. R. China (e-mail: lubai@sdu.edu.cn).}
\thanks{X.~Cheng are with the State Key Laboratory of Photonics and Communications, School of Electronics, Peking University, Beijing, 100871, P. R. China (email: xiangcheng@pku.edu.cn).}
}

% The paper headers
\markboth{IEEE Transactions on Intelligent Transportation Systems, vol. xx, no. xx, XX 2026}
{Submitted paper}
% \IEEEpubid{0000--0000/00\$00.00~\copyright~2021 IEEE}
% Remember, if you use this you must call \IEEEpubidadjcol in the second
% column for its text to clear the IEEEpubid mark.

\maketitle
\begin{abstract}
To support sixth-generation (6G)-enabled intelligent transportation systems (ITSs), a multi-modal sensing residual-corrected graph neural network (MM-ResGNN) framework is proposed for millimeter-wave (mmWave) path loss prediction in vehicular communications for the first time. The propagation environment is formulated as an environment sensing path loss graph (ESPL-Graph), where nodes represent the transmitter (Tx) and receiver (Rx) entities and edges jointly describe Tx--Rx transmission links and Rx--Rx spatial correlation links. Meanwhile, a geometry-driven physical baseline is introduced to decouple deterministic attenuation trends from stochastic residual variations. A vehicular multi-modal path loss dataset (VMMPL) is constructed, which covers three representative scenarios, including the urban wide lane, urban crossroad, and suburban forking road environments, and achieves precise alignment between RGB images and global semantic information in the physical space, and link-level ray-tracing (RT)-based path loss data in the electromagnetic space. In MM-ResGNN, topology-aware graph representations and fine-grained visual semantics are synergistically integrated through a gated fusion mechanism to estimate the path loss residual relative to the physical baseline. Experimental results demonstrate that MM-ResGNN achieves significant improvements over empirical models and conventional data-driven baselines, with a normalized mean squared error (NMSE) of 0.0098, a mean absolute error (MAE) of 5.7991~dB, and a mean absolute percentage error (MAPE) of 5.0498\%. Furthermore, MM-ResGNN exhibits robust cross-scenario generalization through a few-shot fine-tuning strategy, enabling accurate path loss prediction in unseen vehicular environments with limited labeled data.
\end{abstract}

%投稿前检查关键词要求
\begin{IEEEkeywords}
Sixth-generation (6G), intelligent transportation systems (ITSs), vehicular networks, path loss prediction, graph neural network (GNN).
\end{IEEEkeywords}

\section{Introduction}
Intelligent transportation systems (ITSs) constitute a core component of modern smart cities and play a critical role in traffic safety enhancement, mobility improvement, and sustainable transportation development \cite{ITS1,6G_V2X_Review,6G_V2X_Slicing}. 
With the evolution toward sixth-generation (6G)-enabled ITSs, reliable vehicular connectivity requires accurate channel characterization to support high-mobility communications and stringent quality-of-service requirements \cite{6G_Enabled_ATS,channel_1}. 
Path loss plays a fundamental role in quantifying signal attenuation and serves as a cornerstone for network coverage planning, power control, and interference management \cite{PL_application1,PL_application2,PL_application3}. 
In realistic vehicular environments, signal propagation is jointly affected by static elements, such as buildings and roadside infrastructure, and dynamic elements, such as vehicles, forming a heterogeneous propagation space with complex interactions. 
Consequently, path loss depends not only on individual transmitter (Tx) and receiver (Rx) characteristics, but also exhibits pronounced spatial dependency and topological correlation among multiple communication links \cite{PL_tx_rx}, which substantially increases the difficulty of accurate prediction.

Path loss modeling has been extensively investigated from early-generation cellular systems to contemporary fifth-generation (5G) networks. 
Conventional analytical and empirical path loss models, such as the free-space model and the close-in model, have been widely adopted owing to concise mathematical formulations and physical interpretability \cite{CI_model}. 
Empirical path loss models provide a fundamental characterization of large-scale signal attenuation under simplified propagation assumptions and are typically calibrated utilizing empirically tuned parameters. 
However, the simplified propagation assumptions adopted in the empirical models are insufficient to characterize complex vehicular environments, in which signal propagation is jointly affected by buildings, vehicles, and surrounding infrastructure \cite{empirical_model2,empirical_model3}. 
As a result, limited prediction accuracy and generalization capability are observed when the empirical models are applied to realistic ITS scenarios.

To address these challenges, data-driven approaches have been explored for path loss prediction. Machine learning (ML)-based methods were employed to model nonlinear relationships between environmental features and path loss  in vehicular scenarios \cite{PathLoss_Urban_ML,PathLoss_Manhattan,PathLoss_RT_Sionna}. Conventional ML algorithms, including support vector machines (SVMs) and random forests, demonstrated improved prediction performance compared with empirical models \cite{SVM_RF_INDOOR,V2V_SVM_RF}. However, the conventional ML algorithms typically treated each Tx--Rx link as an independent sample and relied heavily on handcrafted features, thereby neglecting spatial correlation and inter-link dependency in dense road networks. More recently, deep learning (DL)-based approaches leveraged visual representations, such as satellite images and city maps, to infer path loss by extracting high-level semantic information related to building density and road layouts \cite{CNN_Satellite,CNN_City_Map,SatImage_PL_Prediction}. Despite improved generalization, the models based on visual images were generally built on regular grid representations and lacked explicit mechanisms to capture network topology and link-level interactions.

Graph neural networks (GNNs) have been recognized as an effective tool for modeling structured wireless systems by explicitly representing network entities and their interactions. 
GNN-based approaches have demonstrated promising performance in various wireless applications, including resource allocation, routing optimization, and distributed inference \cite{GNN_IoT,GNN_EdgeUpdate,GNN_Routing,GNN_OAC}. 
By aggregating information over network topology, GNNs are well suited to capture spatial dependency and inter-link correlation in wireless networks. 
Nevertheless, the application of GNNs to millimeter-wave (mmWave) path loss prediction remains limited. 
In existing studies, graph constructions are often simplified and mainly based on geographical proximity or basic link geometry \cite{GNN_PathLoss_Geo,GNN_ChannelMod}. 
Such simplified graph constructions lead to low-dimensional representations, which are insufficient to characterize complex blockage patterns and link heterogeneity in realistic vehicular environments. 

In realistic mmWave vehicular environments, signal propagation is highly sensitive to environmental characteristics, including obstacle types, local structural configurations, and the presence of dynamic objects. 
Environmental sensing modalities, such as visual images, are of great importance for capturing fine-grained semantic information that is closely related to mmWave propagation mechanisms. 
However, such environmental sensing information is rarely incorporated into existing graph-based path loss modeling frameworks. 
As a result, fine-grained semantic information related to obstacle types, local structural configurations, and dynamic objects is difficult to be adequately represented when path loss prediction relies solely on geometric features. 

Inspired by synesthesia of machines (SoM) \cite{SoM} and multi-modal intelligent channel modeling (MMICM) \cite{MMICM, WCM}, environmental sensing can be jointly exploited with topology-aware representations to improve path loss prediction in complex vehicular environments. Visual information, such as RGB images, provides semantic cues about obstacles and local structures that are not captured by geometry-only features. However, existing graph-based path loss models rarely incorporate aligned environmental sensing, and publicly available vehicular datasets with synchronized sensing data and link-level path loss labels remain limited. To address these issues, a multi-modal sensing residual-corrected graph neural network (MM-ResGNN) is proposed for mmWave path loss prediction for ITSs. The propagation environment is abstracted as an environment sensing path loss graph (ESPL-Graph), which enables explicit modeling of spatial dependency and inter-link correlation among multiple communication links in dense vehicular networks.
A geometry-driven physical baseline is incorporated to separate deterministic attenuation from residual variations induced by complex environments, while visual semantic information is jointly exploited to capture fine-grained environmental characteristics that are difficult to represent utilizing geometry-only features.
In addition, a vehicular multi-modal path loss (VMMPL) dataset is constructed to support the evaluation of the proposed MM-ResGNN, which achieves precise alignment between RGB images and global semantic information in the physical space, and link-level path loss data in the electromagnetic space across three representative vehicular scenarios. The main contributions are summarized as follows.

\begin{enumerate}
    \item A novel MM-ResGNN framework is developed for vehicular mmWave path loss prediction. The proposed residual-corrected modeling paradigm integrates a geometry-driven physical baseline with multi-modal graph learning to characterize large-scale signal attenuation. The decoupling of deterministic propagation trends and stochastic environmental variations enables accurate and robust path loss estimation in complex intelligent transportation scenarios.

    \item To support the evaluation of the proposed MM-ResGNN, a new VMMPL dataset is constructed, which also facilitates a broad range of future research on multi-modal data-driven path loss modeling, topology-aware learning, and environment-aware mmWave channel characterization for ITSs. The VMMPL dataset achieves precise alignment between RGB images and global semantic information in the physical space, and link-level path loss data in the electromagnetic space and covers three representative vehicular scenarios. The VMMPL dataset operates at a carrier frequency of 28~GHz with a bandwidth of 2~GHz and contains 41.8k synchronized RGB images and corresponding path loss measurements.

    \item To explicitly model spatial dependency and inter-link correlation in vehicular communication environments for ITSs, a novel ESPL-Graph is formulated, which provides a general and extensible graph-based abstraction for representing topology-aware interactions among communication links and can be readily applied to graph learning–based wireless modeling and inference tasks. In the proposed ESPL-Graph formulation, nodes represent Tx and Rx entities, while edges jointly characterize Tx--Rx transmission links and spatial correlation relationships among Rxs. Based on ESPL-Graph, a dedicated MM-ResGNN architecture integrates topology-aware graph representations and fine-grained visual semantics through a gated fusion mechanism for residual path loss estimation.

    \item Experimental results demonstrate that MM-ResGNN achieves significant improvements over empirical models and conventional data-driven baselines, with a normalized mean squared error (NMSE) of 0.0098, a mean absolute error (MAE) of 5.7991~dB, and a mean absolute percentage error (MAPE) of 5.0498\%. The proposed framework further exhibits strong generalization capability across different vehicular scenarios under a few-shot fine-tuning setting, which validates its robustness and practical applicability.
\end{enumerate}

The remainder of this paper is organized as follows. Section~II presents the construction of the VMMPL dataset for mmWave path loss prediction. Section~III describes the architecture and design of the proposed MM-ResGNN. In Section~IV, the simulation settings and a comprehensive performance evaluation and analysis are introduced. Finally, conclusions are drawn in Section~V.

\begin{figure}[!t]
	\centering
	\includegraphics[width=0.48\textwidth]{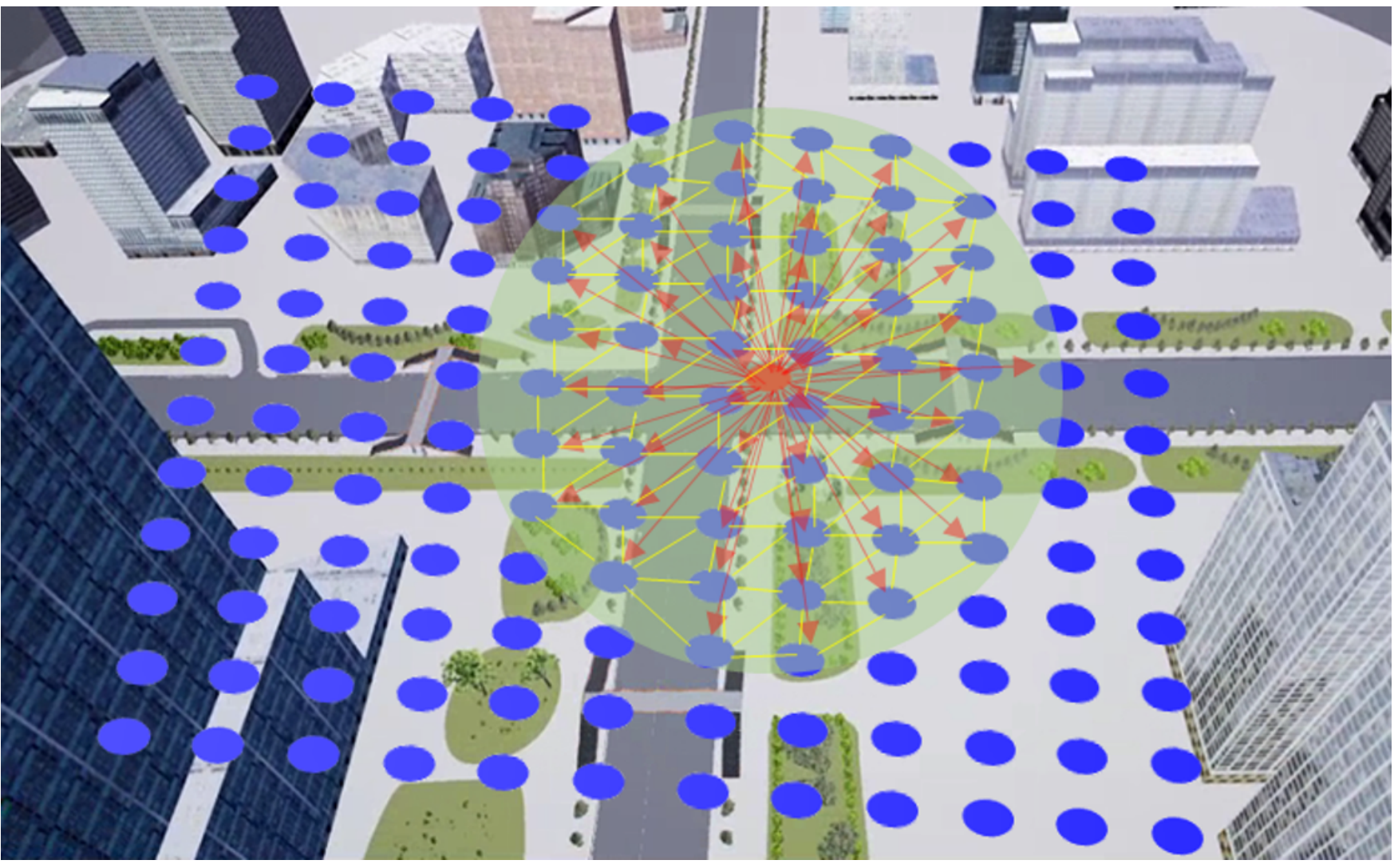}
	\caption{Illustration of the mmWave path loss prediction task in the urban scenario. The Tx (red vehicle) communicates with a dense grid of ground Rxs (blue dots). The prediction targets are the links connected to the nearest 50 Rxs (highlighted in the green zone).}
	\label{pl_task}
\end{figure}

\section{VMMPL Dataset Construction}

To support the development of MM-ResGNN, a vehicular multi-modal path loss dataset, referred to as the VMMPL dataset, is constructed. 
The VMMPL dataset features precise spatio-temporal alignment between RGB images and global semantic information in the physical space, and link-level path loss data in the electromagnetic space.
Dataset construction is based on a heterogeneous co-simulation platform integrating AirSim \cite{AirSim} and Wireless InSite \cite{WI}, where photorealistic physical environments and high-fidelity ray-tracing (RT)-based radio propagation are jointly captured.

As illustrated in Fig.~\ref{pl_task}, the prediction task focuses on estimating the path loss for the $K=50$ nearest Rxs associated with a dynamic vehicular Tx, where dominant communication links in the local vicinity are captured. 
Although $K=50$ is adopted for evaluation, the proposed framework can be readily extended to predict path loss for an arbitrary number of Rxs according to different network configurations and application requirements. In this work, $K$ is fixed to ensure a fair and consistent comparison across scenarios and baselines, while controlling the graph size and computational cost.

\subsection{Scenario Configuration}
\begin{figure}[!t]
	\centering
	\includegraphics[width=0.49\textwidth]{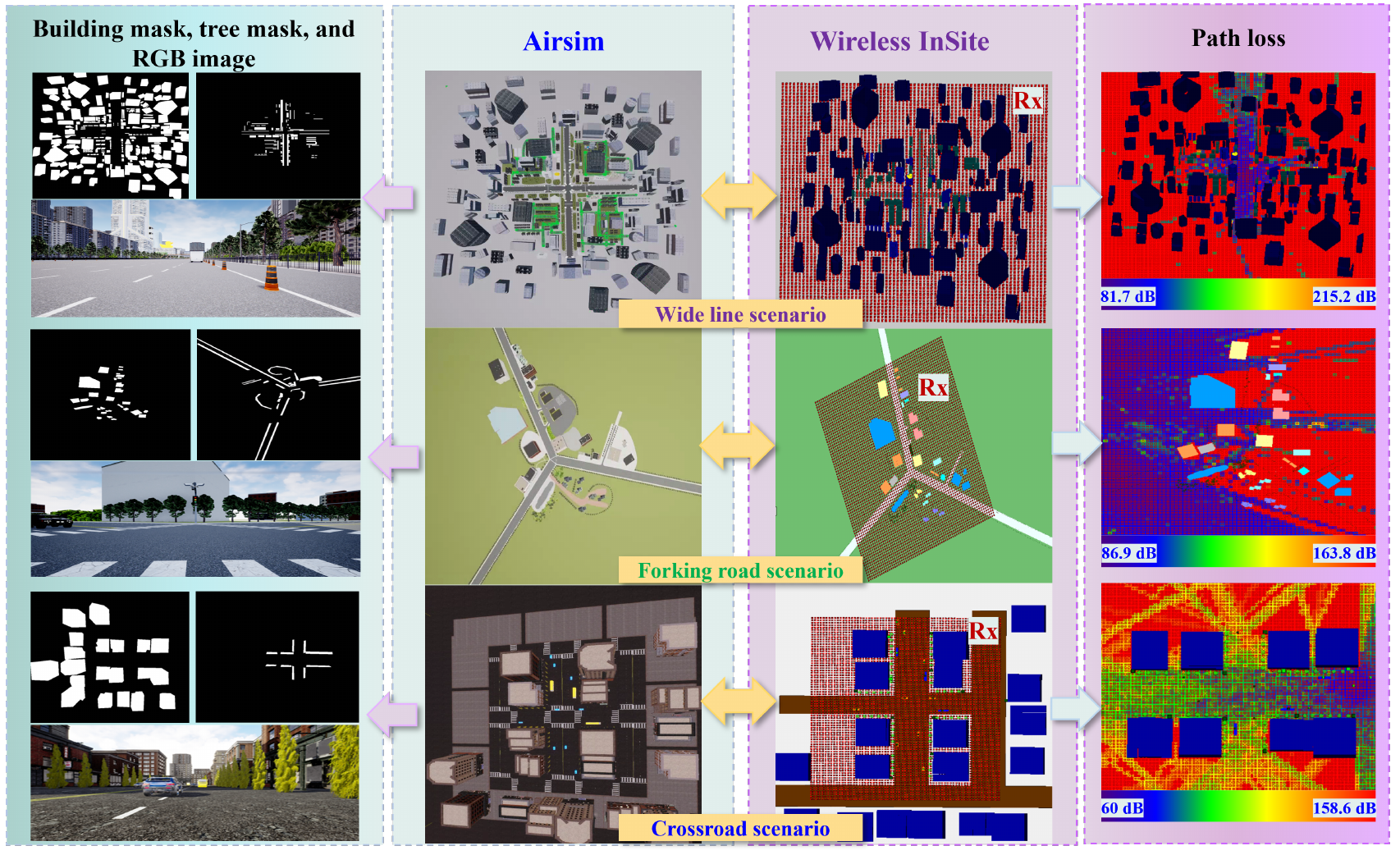}
	\caption{Visualization of the constructed VMMPL dataset across three scenarios. The VMMPL dataset includes global masks (building and tree), ego-centric RGB images captured by the vehicle, aligned 3D physical environments in AirSim and Wireless InSite, and the corresponding ground-truth path loss heatmaps.}
	\label{fig:data_sample}
\end{figure}

The VMMPL dataset comprises three representative road scenarios spanning dense urban to suburban vehicular environments, as illustrated in Fig.~\ref{fig:data_sample}. 
To ensure consistency between AirSim and Wireless InSite, a dual-space alignment mechanism is employed during dataset construction. 
In the geometric domain, three-dimensional models of buildings, roads, and vehicles are synchronized in scale and coordinate origin. 
In the temporal domain, the states of all dynamic elements are updated synchronously at each snapshot to maintain temporal alignment between the physical and electromagnetic spaces.

\begin{figure*}[!t]
	\centering
	\subfloat[Urban Wide Lane]{\includegraphics[width=0.32\textwidth,height=0.28\textwidth]{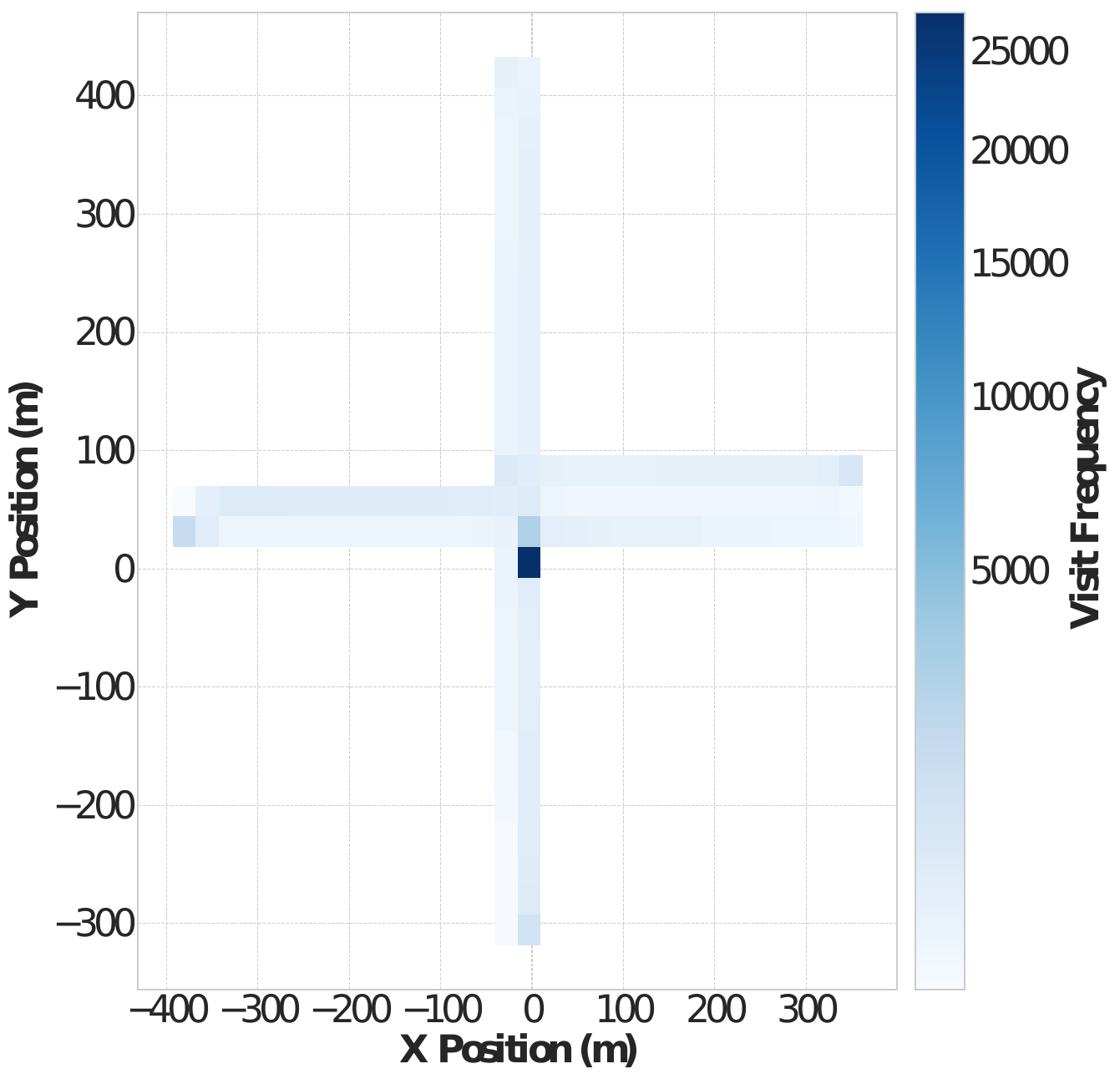}}
    \subfloat[Suburban Forking Road]{\includegraphics[width=0.32\textwidth,height=0.28\textwidth]{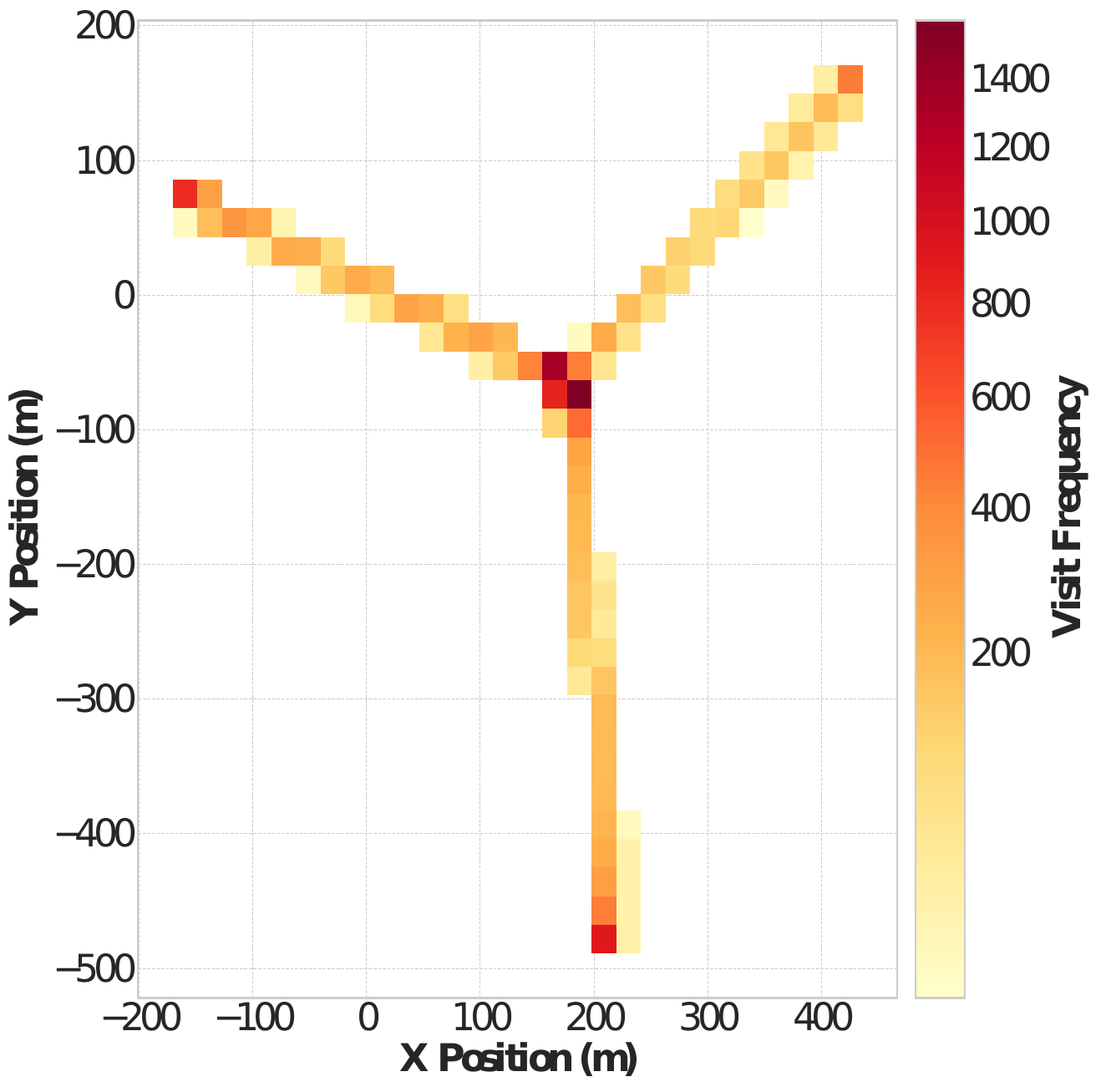}}
	\subfloat[Urban Crossroad]{\includegraphics[width=0.32\textwidth,height=0.28\textwidth]{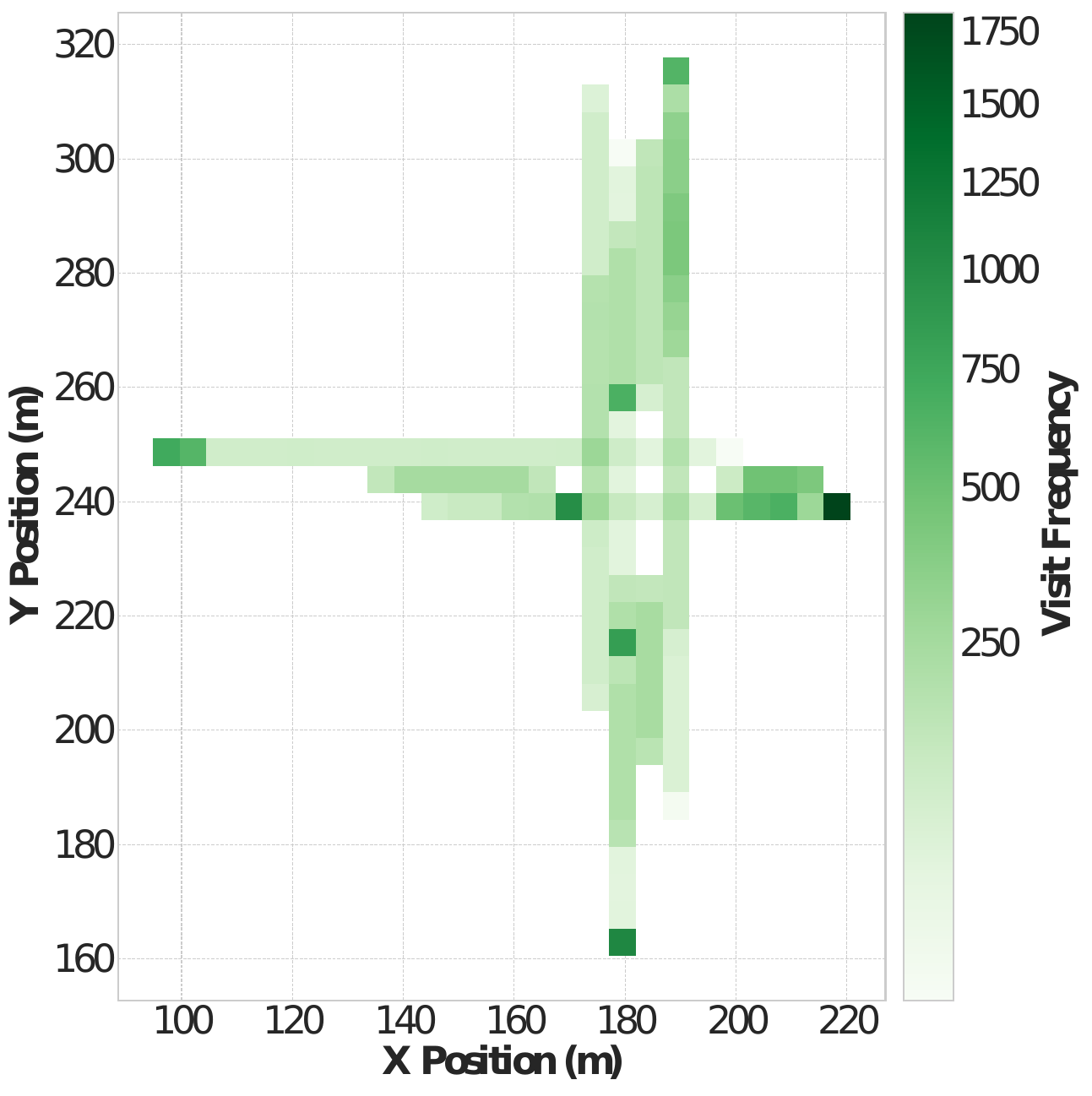}}
	\caption{Vehicle trajectory density heatmaps across the three representative scenarios. The heatmaps illustrate the spatial coverage and diverse mobility patterns, reflecting different traffic densities and roadway topologies.}
	\label{fig:trace_heatmaps}
\end{figure*}

\textbf{Urban Wide Lane (Dense Urban Scenario)}:
The urban wide lane scenario is modeled after a major metropolitan arterial road with an ultra-wide roadway and dense high-rise buildings. 
40 dynamic vehicles are deployed along parallel lanes, and the trajectory density is shown in Fig.~\ref{fig:trace_heatmaps}(a). 
Owing to its large spatial coverage and diverse blockage conditions, the urban wide lane scenario is adopted as the primary training dataset.

\textbf{Suburban Forking Road (Suburban Scenario):}
The suburban forking road scenario represents a Y-shaped road bifurcation located in a low-density environment with sparse low-rise buildings and vegetation coverage. 
25 dynamic vehicles are deployed along the bifurcated road segments, and the trajectory density is shown in Fig.~\ref{fig:trace_heatmaps}(b). 
The suburban forking road scenario is utilized to examine model adaptability in open suburban vehicular environments.

\textbf{Urban Crossroad (Regular Urban Scenario):}
The urban crossroad scenario corresponds to a bidirectional intersection surrounded by medium-rise buildings arranged in a grid-based Manhattan layout. 
20 dynamic vehicles are deployed within the intersection area, and the trajectory density is shown in Fig.~\ref{fig:trace_heatmaps}(c). 
The urban crossroad scenario is employed to evaluate cross-scenario generalization under regular urban topologies.

\subsection{Data Collection and Processing}

After scenario construction and platform synchronization, a data generation pipeline is established to deploy communication transceivers, record synchronized multi-modal observations, and convert simulation outputs into graph-structured samples.

\subsubsection{Transceiver Deployment and Trajectory Configuration}

The communication system operates at a carrier frequency of 28~GHz with a bandwidth of 2~GHz. 
The Tx is mounted on dynamic vehicles at a height of 2.0~m, and vehicle trajectories are configured along the road network. 
A dense grid of Rxs is deployed on the ground plane at a height of 1.5~m, including 5{,}183 nodes in the urban wide lane scenario, 5{,}159 nodes in the urban crossroad scenario, and 4{,}964 nodes in the suburban forking road scenario.  
The resulting VMMPL dataset contains 32.3k snapshots $\times$ 5.18k links per snapshot for the wide lane scenario, 4.42k snapshots $\times$ 5.16k links per snapshot for the crossroad scenario, and 5.07k snapshots $\times$ 4.96k links per snapshot for the forking road scenario.

\begin{figure*}[!t]
	\centering
	\includegraphics[width=0.95\textwidth]{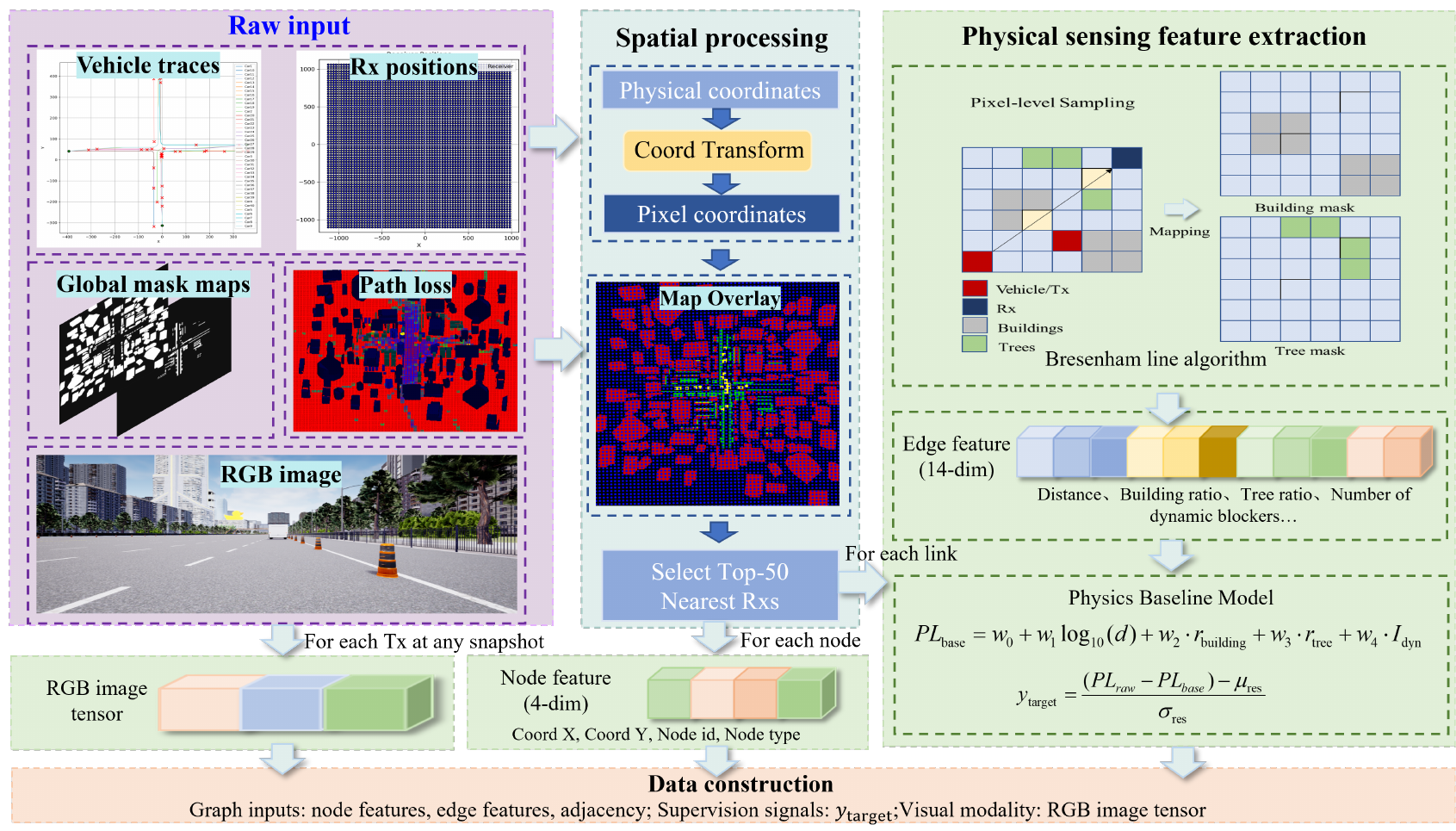}
	\caption{Workflow of the data processing and graph construction pipeline. The raw multi-modal data (left) are spatially aligned and transformed into pixel coordinates. A physics-aware feature extraction module (right) utilizes the Bresenham line algorithm to compute blockage features from semantic masks. Finally, a physical baseline model is applied to decouple the deterministic path loss component, generating the graph-structured inputs and residual targets for the neural network.}
	\label{dataset}
\end{figure*}

\subsubsection{Multi-Modal Data Acquisition}

For each snapshot, multi-modal data are recorded in a synchronized manner, as illustrated in Fig.~\ref{fig:data_sample}. 
The visual modality includes ego-centric RGB images captured from the Tx perspective. 
In addition, global binary masks for buildings and trees are generated to support blockage-related analysis. 
The electromagnetic modality consists of ground-truth path loss values for all Tx--Rx links computed utilizing RT in Wireless InSite. 
Physical state information is also logged, including the global 3D coordinates of transceivers and other dynamic vehicles, which supports geometry-related feature extraction.

\subsubsection{ESPL-Graph Formulation and Residual Target Construction}

To transform the VMMPL dataset into graph-structured representations compatible with MM-ResGNN, an ESPL-Graph is formulated for each Tx, as illustrated in Fig.~\ref{dataset}. 
For a given Tx, a local communication graph is defined as $\mathcal{G} = (\mathcal{V}, \mathcal{E})$, where the node set $\mathcal{V}$ consists of the Tx and its $K=50$ nearest Rxs. The edge set $\mathcal{E}$ is composed of two types of edges. 
Transmission edges connect the Tx to each associated Rx to represent Tx--Rx links. 
Correlation edges connect neighboring Rxs to model spatial correlation induced by common blockage conditions and surrounding structures. 
The correlation edges are constructed based on a $k$-nearest neighbor (KNN) criterion in the spatial domain \cite{KNN}. 

Each node in $\mathcal{V}$ is initialized with a feature vector comprising its spatial coordinates and transceiver type, i.e., indicating whether the node is a Tx or an Rx. For each Tx--Rx transmission edge, a 14-dimensional physics-aware feature vector is extracted to describe geometric and environmental attributes along the propagation path. 
Feature extraction is performed utilizing the Bresenham ray-casting algorithm, as shown in the right panel of Fig.~\ref{dataset}. 
The extracted features include the logarithmic Tx--Rx distance, blockage ratios associated with buildings and trees, and normalized distances to dynamic blockers such as vehicles. 
The feature vector is assigned as the edge attribute of the corresponding transmission Tx--Rx edge. Correlation edges connect neighboring Rxs to facilitate the exchange of node information, modeling spatial correlation induced by common blockage conditions and surrounding structures.

To decouple large-scale deterministic attenuation trends from complex stochastic variations, a geometry-driven physical baseline model is adopted for Tx--Rx transmission edges as a fixed reference for residual learning. 
The physical baseline path loss is defined as
$
PL_{base} = w_0 + w_1 \log_{10}(d) + w_2 \cdot r_{\text{building}} + w_3 \cdot r_{\text{tree}} + w_4 \cdot I_{\text{dyn}}
$
where $d$ denotes the Tx--Rx distance, $r_{\text{building}}$ and $r_{\text{tree}}$ represent the blockage ratios of buildings and trees along the propagation path, respectively, and $I_{\text{dyn}}$ indicates the presence of dynamic blockers. 
The coefficients $\{w_i\}$ are pre-estimated via least-squares fitting on the training subset of the VMMPL dataset and remain fixed during model training and inference. 
Based on the physical baseline, the raw path loss value associated with each Tx--Rx transmission edge is transformed into a normalized residual learning target given by
$
y_{\text{target}} = \frac{(PL_{raw} - PL_{base}) - \mu_{\text{res}}}{\sigma_{\text{res}}}
$
where $\mu_{\text{res}}$ and $\sigma_{\text{res}}$ denote the mean and standard deviation of the residuals computed on the training set.

\subsection{Statistical Analysis and Physical Insights}

\begin{figure}[!t]
	\centering
	\subfloat[Wide Lane scenario]{\includegraphics[width=0.49\textwidth]{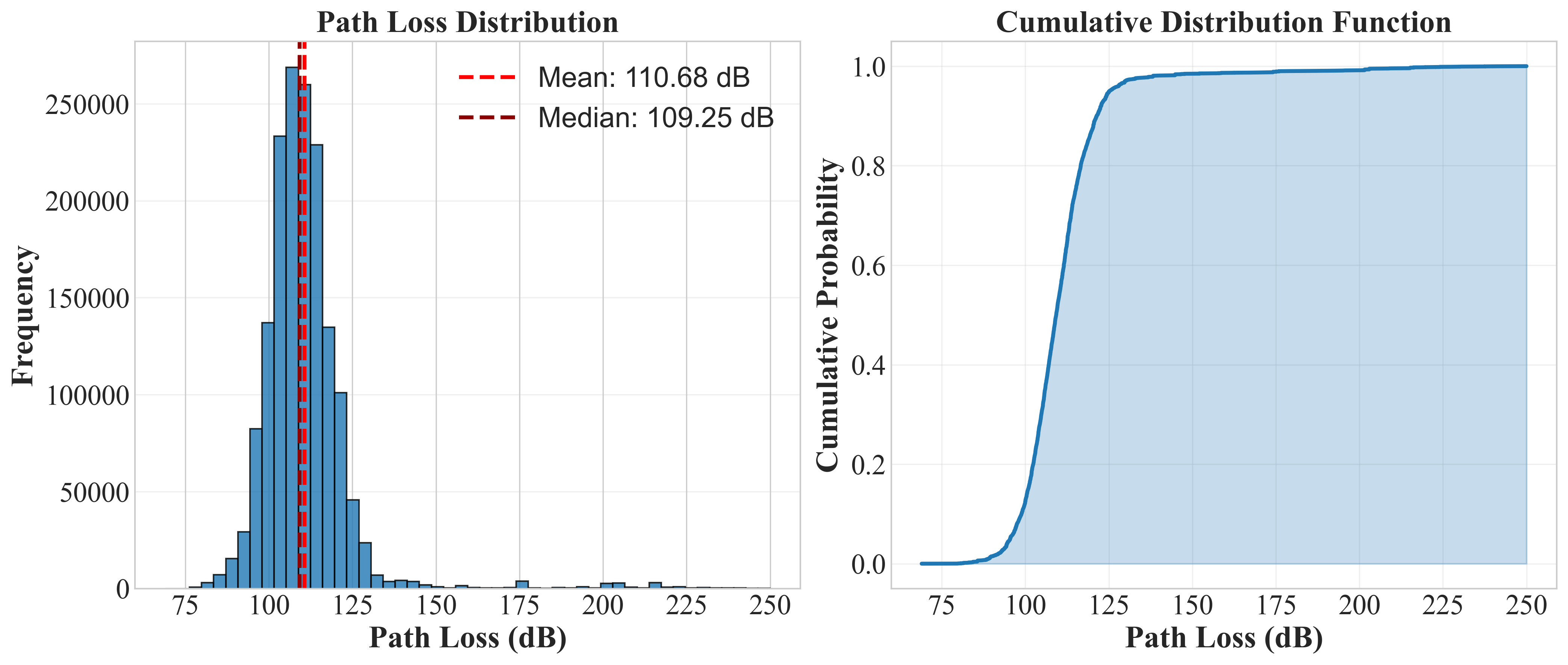}}
	\\
	\subfloat[Forking Road scenario]{\includegraphics[width=0.49\textwidth]{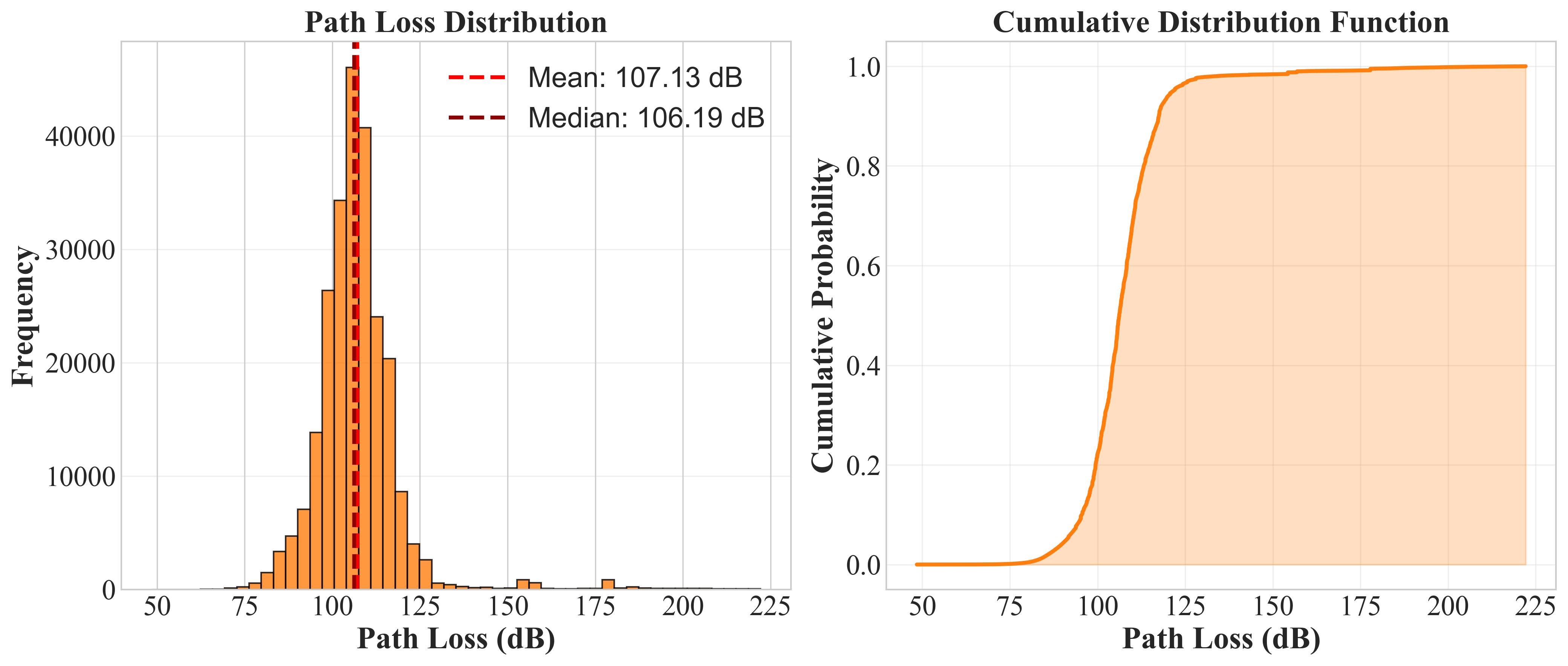}}
    \\
	\subfloat[Crossroad scenario]{\includegraphics[width=0.49\textwidth]{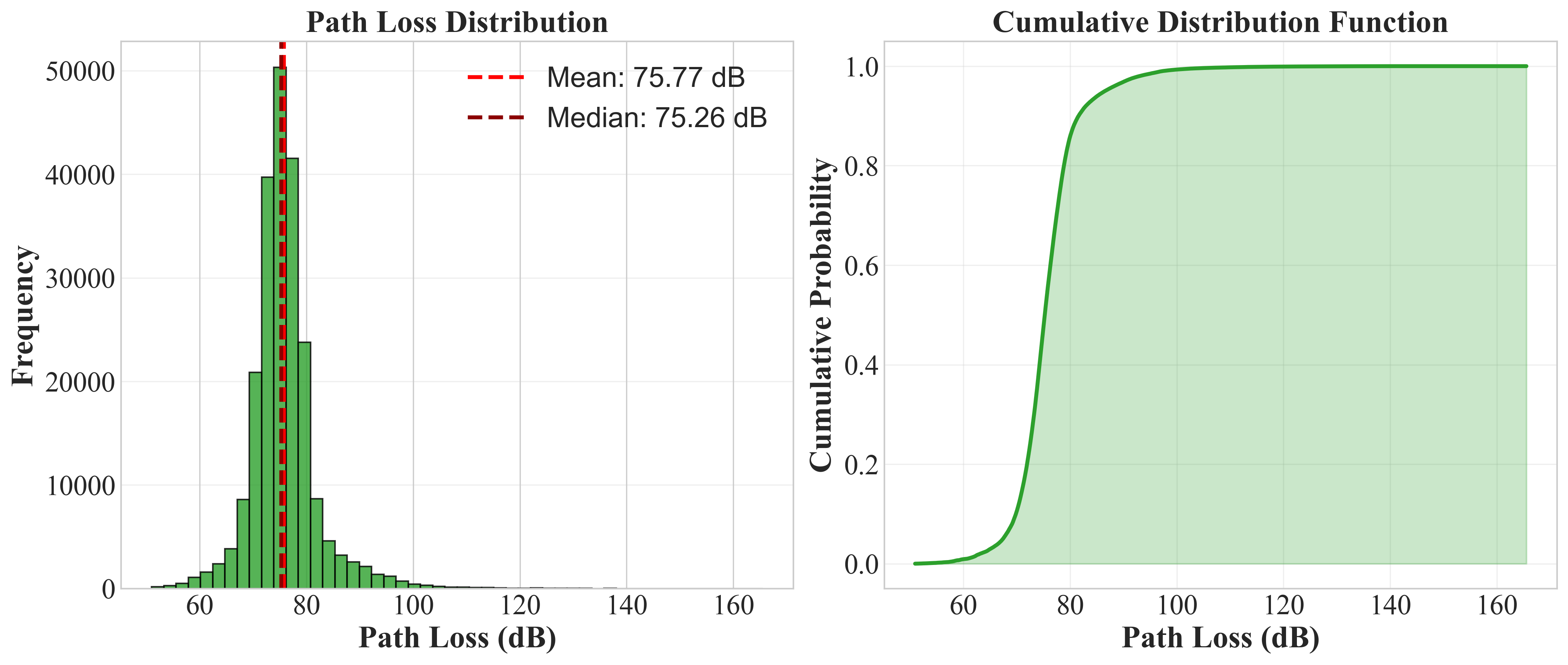}}
	\caption{Statistical distribution of the path loss across three scenarios. The left panels present the frequency histograms with mean and median markers, while the right panels depict the empirical CDFs, highlighting the distinct electromagnetic scales.}
	\label{fig:pl_stats}
\end{figure}

To examine the diversity of the constructed VMMPL dataset, path loss statistics are analyzed for the three scenarios. 
The analysis is conducted based on the Tx--Rx links formed between each Tx and its nearest 50 Rxs, and the histograms and empirical cumulative distribution functions (CDFs) are shown in Fig.~\ref{fig:pl_stats}.

The urban wide lane scenario exhibits a wide path loss dynamic range with a mean value of 110.68~dB, as illustrated in Fig.~\ref{fig:pl_stats}(a). 
High-rise buildings and large-scale blockages produce a tail in the high-loss region above 150~dB and reflect deep fading conditions in non-line-of-sight (NLoS) areas. 
The suburban forking road scenario presents a mean path loss of 107.13~dB, as depicted in Fig.~\ref{fig:pl_stats}(b). 
The urban crossroad scenario yields a lower mean path loss of 75.77~dB, as shown in Fig.~\ref{fig:pl_stats}(c). 
The separation among these statistical profiles confirms that the constructed VMMPL dataset captures distinct propagation characteristics across diverse intelligent transportation system environments, which provides a comprehensive benchmark for evaluating prediction accuracy and cross-scenario generalization capability of the proposed MM-ResGNN.

\section{Framework of the Proposed MM-ResGNN}
\begin{figure*}[!t]
\centering
\includegraphics[width=\textwidth]{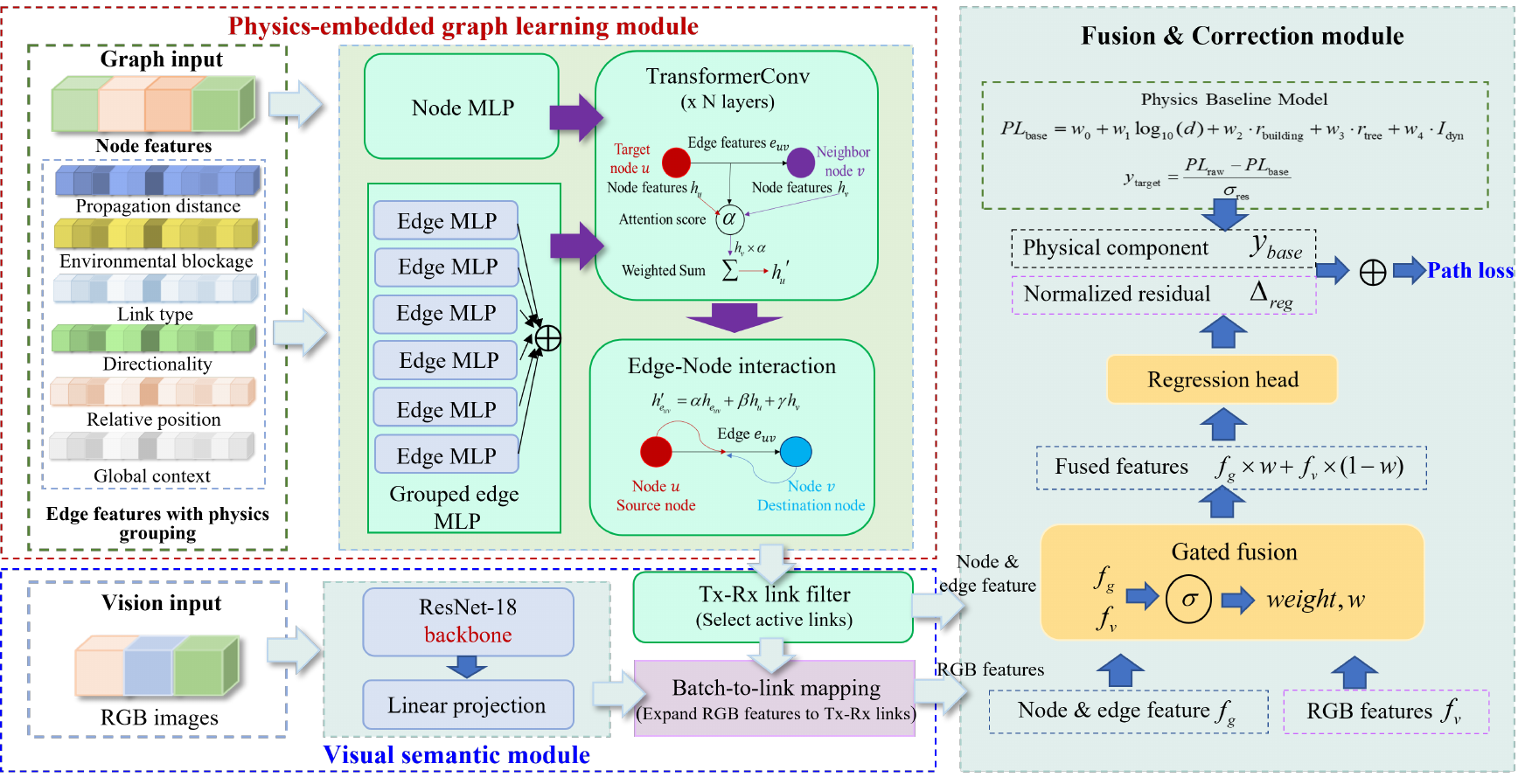}
\caption{The overall architecture of MM-ResGNN. The framework comprises two parallel branches, i.e., a physics-embedded graph learning branch for topological reasoning and a visual semantic branch for environmental texture extraction. A gated fusion module dynamically integrates these multi-modal features to predict the normalized residual, which is added to the physical baseline to obtain the final path loss.}
\label{model}
\end{figure*}

In this section, the overall framework of the proposed MM-ResGNN is presented. 
As illustrated in Fig.~\ref{model}, the proposed architecture follows a dual-branch residual correction paradigm for mmWave path loss prediction in complex vehicular environments. 
The upper left module in Fig.~\ref{model} corresponds to a physics-embedded graph learning module, which takes graph-structured inputs with physics-aware node and edge features and performs topology-aware message passing to model spatial dependency and inter-link correlation. 
In parallel, the bottom left module in Fig.~\ref{model} represents the visual semantic module, which extracts high-level environmental semantics from ego-centric RGB images and maps them to Tx--Rx links. 
The outputs of these two branches are fused through a gated cross-modal fusion and correction module, shown on the right side of Fig.~\ref{model}, which adaptively balances geometry-driven and visual features to predict a normalized residual. 
The predicted residual is finally combined with a geometry-driven physical baseline to obtain the final path loss estimate.

\subsection{Physics-Embedded Graph Learning Branch}
To capture the structural dependency of wireless channels, a physics-embedded graph learning branch is designed, corresponding to the upper left module in Fig.~\ref{model}. 
This module operates on the ESPL-Graph and processes physics-aware node and edge features derived from geometric and environmental attributes. 
Unlike conventional GNNs that treat features as abstract vectors, this module explicitly embeds domain knowledge into feature encoding and message passing. 
A physics-aware grouped encoding strategy and an anisotropic aggregation mechanism are jointly employed to ensure that the learned representations are both physically meaningful and topologically consistent.

\subsubsection{Physics-Aware Grouped Edge Encoding}
As shown in the upper left module of Fig.~\ref{model}, edge features with physics grouping are processed by a physics-aware grouped encoding module. 
Directly projecting heterogeneous edge features into a shared latent space may suppress contributions from smaller-scale physical factors, such as environmental blockages. 
To mitigate this issue, the 14-dimensional edge feature vector is partitioned into six sub-groups according to their physical meanings, including propagation distance, environmental blockage, link type, and relative position. 
Each sub-group is processed by an independent edge multilayer perceptron (MLP), and the resulting embeddings are concatenated and fused through a linear projection layer followed by layer normalization.

\subsubsection{Anisotropic Topological Aggregation via Graph Transformer}
The encoded node and edge features are then fed into a stack of TransformerConv layers, as illustrated in the central part of Fig.~\ref{model}. 
By leveraging multi-head attention, the graph transformer enables anisotropic message aggregation, which is essential for modeling directional signal propagation and blockage-sensitive mmWave channels \cite{GTN_NeurIPS19}.

For a center node $u$ and neighboring node $v$, the feature update at layer $l$ is expressed as
\begin{equation}
    \mathbf{h}_u^{(l+1)} = \mathbf{W}_O \cdot \text{Concat}_{k=1}^{H}
    \left(
        \sum_{v \in \mathcal{N}(u)}
        \alpha_{uv}^{(k)} \mathbf{W}_V^{(k)} \mathbf{h}_v^{(l)}
    \right)
\end{equation}
where $\mathcal{N}(u)$ denotes the neighborhood set of node $u$, and $H$ is the number of attention heads. The attention coefficient $\alpha_{uv}^{(k)}$ measures the importance of neighboring node $v$ to node $u$ under the $k$-th attention head, computed as
\begin{equation}
    \alpha_{uv}^{(k)} =
    \text{Softmax}_{v}
    \left(
    \frac{
        (\mathbf{W}_Q^{(k)} \mathbf{h}_u^{(l)})^T
        \big(
            \mathbf{W}_K^{(k)} \mathbf{h}_v^{(l)} +
            \mathbf{W}_E^{(k)} \mathbf{h}_{e_{uv}}^{(l)}
        \big)
    }
    {\sqrt{d_k}}
    \right)
\end{equation}
where $\mathbf{W}_Q^{(k)}$, $\mathbf{W}_K^{(k)}$, and $\mathbf{W}_E^{(k)}$ are the projection matrices for queries, keys, and edge attributes, respectively.

An adaptive edge–node fusion mechanism is utilized to refine the edge representation after message passing. For a directed edge $e_{uv}$, the updated edge feature is computed as
\begin{equation}
    \mathbf{h}_{e_{uv}}^{(l+1)} =
    \alpha \cdot \mathbf{h}_{e_{uv}}^{(l)} +
    \beta \cdot \mathbf{h}_{u}^{(l)} +
    \gamma \cdot \mathbf{h}_{v}^{(l)}
\end{equation}
where $\alpha$, $\beta$, and $\gamma$ are learnable fusion coefficients, obtained via a softmax-based gating network.

\subsection{Visual Semantic Branch}
The visual semantic branch, shown in the bottom left module of Fig.~\ref{model}, aims to incorporate fine-grained environmental semantics from ego-centric RGB images into path loss prediction. 
This branch is composed of two closely related components. 
First, high-level semantic representations are extracted from RGB images to characterize local environmental textures and structures. 
Then, the extracted visual features are mapped to graph-structured Tx--Rx links in a topology-aware manner, such that visual semantics are associated only with transmission links relevant to path loss estimation. 
Based on this two-stage design, visual information is effectively aligned with the graph-based geometric representations.

\subsubsection{Semantic Feature Extraction and Transfer Learning}
A pre-trained ResNet-18 \cite{Resnet} is utilized to extract high-level semantic representations. Transfer learning is employed to fine-tune the network for path loss prediction. Shallow layers are frozen, preserving universal spatial features, while high-level layers are fine-tuned to extract domain-specific features related to path loss in vehicular environments.

\subsubsection{Topology-Aware Feature Mapping}
As illustrated by the Tx--Rx link filter in Fig.~\ref{model}, visual features are selectively associated with Tx--Rx transmission edges, while Rx--Rx correlation edges are assigned zero-padding. 
This design ensures that visual semantics enhance path loss estimation only for transmission links without contaminating correlation modeling.

\subsection{Gated Cross-Modal Fusion and Residual Correction}
The gated cross-modal fusion and correction module, shown on the right side of Fig.~\ref{model}, adaptively integrates geometry-driven graph features and visual semantic features. 
A gating coefficient is learned to balance the contributions of the two modalities under different propagation conditions. 
The fused representation is utilized to predict a normalized residual, which is added to the geometry-driven physical baseline to obtain the final path loss estimate. The gating coefficient $w \in [0,1]$ is computed as
\begin{equation}
    w = \sigma\left(\mathbf{W}_g \left[\mathbf{f}_{geo} \parallel \mathbf{f}_{vis}\right] + \mathbf{b}_g\right)
\end{equation}
where $\sigma(\cdot)$ is the sigmoid activation function. The final fused representation is
\begin{equation}
    \mathbf{f}_{final} = w \cdot \mathbf{f}_{geo} + (1 - w) \cdot \mathbf{f}_{vis}.
\end{equation}

The fused representation is utilized to predict the normalized residual, which is added to the physical baseline to obtain the final path loss estimate:
\begin{equation}
    \hat{PL}
    =
    PL_{base}(\mathbf{x}_{phy}) + \left(\Delta \hat{PL} \cdot \sigma_{res} + \mu_{res}\right)
\end{equation}
This formulation encourages the network to focus on complex propagation effects caused by local environmental structures and dynamic obstacles, improving prediction accuracy.

\section{Simulation Results and Analysis}
This section presents simulation results for mmWave path loss prediction and provides a mechanism-oriented analysis of the proposed MM-ResGNN. The evaluation includes comparisons with empirical and data-driven baselines, ablation studies on key architectural components, and cross-scenario transfer learning under limited target-domain supervision.

\subsection{Experimental Setup and Implementation Details}

\subsubsection{Experimental Setup}
A vehicle-wise data splitting strategy is adopted to evaluate generalization under unseen mobility patterns. Instead of randomly shuffling snapshots, the dataset is partitioned according to unique vehicle IDs. Specifically, 70\% of the vehicles and their complete trajectories are utilized for training, 15\% for validation, and 15\% for testing. This protocol prevents information leakage caused by temporally correlated snapshots and provides a rigorous assessment of generalization.

The proposed model is implemented in PyTorch and optimized utilizing AdamW with an initial learning rate of $5\times 10^{-4}$ and a weight decay of $5\times 10^{-5}$. Training is conducted for 200 epochs with a batch size of 32, and a plateau-based learning rate scheduler is applied. The GNN branch consists of three TransformerConv layers with four attention heads and a hidden dimension of 128. The visual branch adopts an ImageNet-pretrained ResNet-18 backbone and is fine-tuned following the strategy described in Section~III.

\subsubsection{Evaluation Metrics}
Three metrics are employed to quantify prediction accuracy from complementary perspectives. 
The mean absolute error (MAE) is adopted to measure the average absolute deviation in decibels and is defined as
\begin{equation}
    \text{MAE} = \frac{1}{N} \sum_{i=1}^N |y_i - \hat{y}_i|
\end{equation}
where $y_i$ and $\hat{y}_i$ denote the ground-truth and predicted path loss of the $i$-th link, respectively, and $N$ is the number of evaluated samples. 
MAE directly reflects the link-level prediction error on the dB scale and is therefore particularly suitable for assessing absolute path loss deviations in mmWave systems with large dynamic ranges.

The normalized mean squared error (NMSE) is adopted to evaluate the overall goodness of fit relative to the signal energy and is formulated as
\begin{equation}
    \text{NMSE} = \frac{\sum_{i=1}^N (y_i - \hat{y}_i)^2}{\sum_{i=1}^N y_i^2}.
\end{equation}
By normalizing the squared error by the energy of the path loss values, NMSE enables fair comparison across scenarios with different path loss dynamic ranges.

The mean absolute percentage error (MAPE) is computed to assess relative prediction accuracy and is given by
\begin{equation}
    \text{MAPE} = \frac{1}{N} \sum_{i=1}^N \left| \frac{y_i - \hat{y}_i}{y_i} \right| \times 100\%.
\end{equation}
MAPE emphasizes relative deviations and complements MAE by normalizing errors with respect to the path loss magnitude. Unless otherwise specified, all metrics are computed on the reconstructed path loss $\hat{PL}$ after adding the predicted residual to the physical baseline described in Section~II.

\subsection{Performance Evaluation in the Urban Wide Lane Scenario}

\subsubsection{Baselines}
For quantitative comparison, the proposed MM-ResGNN is evaluated against representative empirical path loss models and data-driven baselines. The selected baselines are designed to isolate the effects of physical modeling, geometric features, graph topology, and visual semantics.
For clarity, the evaluated baselines are grouped according to their modeling paradigms and denoted by category-level identifiers. 
Specifically, C0 denotes the geometry-driven physical baseline, which is given in Section II-B. C1 represents standardized and empirical path loss models, and C2--C4 correspond to representative data-driven baselines with increasing modeling complexity.

\paragraph{Empirical Models (C1)}
To benchmark the proposed MM-ResGNN against authoritative and widely adopted analytical baselines, several representative empirical path loss models are considered. 
These models serve as fundamental references in mmWave channel modeling and 5G standardization, and primarily characterize large-scale propagation behavior under simplified physical assumptions.

Free-Space Path Loss (FSPL)~\cite{CI}:
FSPL represents the theoretical lower bound of path loss under ideal line-of-sight (LoS) conditions without any blockage or scattering and is widely adopted as a fundamental reference model in wireless communications. 
It is given by
\begin{equation}
    PL_{\text{FSPL}}(d) = 32.4 + 20 \log_{10}(d) + 20 \log_{10}(f_c),
\end{equation}
where $d$ is the Tx--Rx distance in meters and $f_c$ is the carrier frequency in GHz.

3GPP TR 38.901 Urban Microcell (UMi)~\cite{3GPP}:
The UMi path loss model follows the standardized formulations specified in 3GPP TR~38.901, which constitutes the authoritative channel modeling reference for 5G new radio (NR) systems. 
This model has been extensively utilized in both academic research and industrial evaluation for UMi mmWave deployments. 
In this work, the corresponding 3GPP expressions and recommended parameter settings are adopted without site-specific calibration.

Alpha-Beta-Gamma (ABG)~\cite{Rappaport2015ABG}:
The ABG model is a measurement-driven empirical path loss model derived from extensive multi-frequency propagation campaigns and has been widely recognized as a benchmark model for mmWave channel characterization. 
It expresses path loss as a function of distance and frequency:
\begin{equation}
    PL_{\text{ABG}}(d) = 10 \alpha \log_{10}(d) + \beta + 10 \gamma \log_{10}(f_c),
\end{equation}
where $\alpha$ controls distance dependence, $\beta$ is an offset term, and $\gamma$ captures frequency dependence.

\paragraph{Data-Driven Baselines}
To systematically evaluate the contribution of each module and the superiority of the proposed architecture, three categories of data-driven baselines are implemented.

Geometry-driven MLP (C2):
A MLP is employed to regress the path loss residual directly from the 14-dimensional physics-aware edge feature vector. Each Tx--Rx link is treated as an independent sample, and neither graph topology nor visual environmental information is incorporated. This MLP baseline reflects the prediction accuracy achievable utilizing localized geometric features only.

Vision-based ResNet-18 (C3):
A convolutional neural network (CNN) based on ResNet-18 is utilized to evaluate the standalone contribution of visual information. Semantic features are extracted from ego-centric RGB images and mapped to path loss values through a regression head. Explicit geometric features, such as propagation distance and blockage ratios, are not included.

Uni-modal geometric GNN (C4):
A graph-based model is constructed utilizing the same heterogeneous graph topology and physics-aware edge features as the proposed MM-ResGNN, while excluding the visual semantic branch. This configuration isolates the contribution of graph topology and geometric reasoning without multi-modal fusion.

\subsubsection{Performance Analysis}
\begin{figure}[!t]
    \centering
    \includegraphics[width=0.48\textwidth]{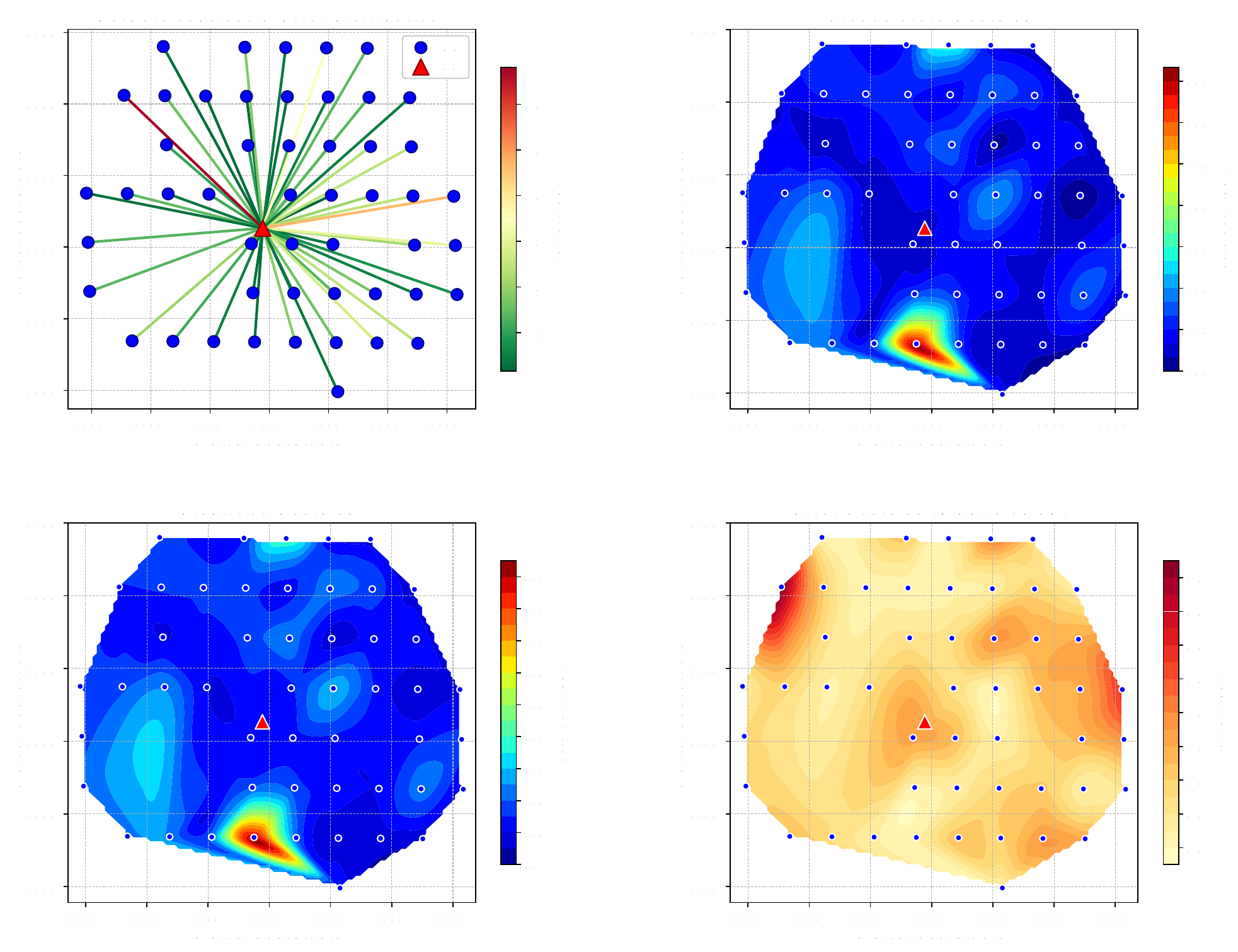}
    \caption{Visualization of path loss prediction results for a representative snapshot in the urban wide lane scenario. The communication graph topology is illustrated with edge colors indicating the prediction error magnitude. The ground-truth and predicted path loss heatmaps are presented together with the spatial distribution of prediction errors. The MAE for the illustrated snapshot is 0.74~dB.}
    \label{fig:frame_viz}
\end{figure}

Before presenting quantitative comparisons, Fig.~\ref{fig:frame_viz} provides a representative snapshot-level visualization in the urban wide lane scenario, which qualitatively illustrates the spatial distribution of path loss and highlights localized prediction errors caused by complex blockage and topology.
Quantitative path loss prediction results in the urban wide lane scenario are summarized in Table~\ref{tab:main_comparison}. The proposed MM-ResGNN is compared with the physical baseline, standardized empirical models, and representative uni-modal data-driven baselines. MAE is primarily utilized to reflect absolute prediction deviations on the dB scale, while MAPE emphasizes relative errors and is more sensitive to inaccuracies in low path loss regions. 
NMSE is further reported to assess the overall goodness of fit normalized by signal energy, providing a complementary global measure that enables fair comparison across scenarios with different path loss dynamic ranges.  

As reported in Table~\ref{tab:main_comparison}, empirical models (C1) exhibit limited accuracy in the urban wide lane scenario. The 3GPP UMi model, ABG model, and FSPL yield MAE values exceeding 11~dB. These results indicate that empirical formulations calibrated on averaged propagation conditions are insufficient to capture site-specific blockage and multipath effects. Compared with standardized empirical models, the geometry-driven physical baseline (C0) achieves a noticeable performance improvement across all evaluation metrics. 
By explicitly incorporating site-aware geometric information, such as propagation distance and blockage ratios, the physical baseline is able to partially capture environment-dependent attenuation effects that are not represented in averaged empirical formulations. 
However, the performance gap between the physical baseline and data-driven models indicates that the low-dimensional parametric formulation of C0 remains insufficient to model complex interactions induced by heterogeneous structures and dynamic obstacles in realistic vehicular environments.

The uni-modal data-driven baselines highlight the limitations of relying on a single modality.
The geometry-driven MLP (C2) and the uni-modal geometric GNN (C4) achieve lower prediction errors than empirical models (C1) and the geometry-driven physical baseline (C0), indicating that geometric features and topological modeling provide an effective inductive bias for capturing large-scale attenuation trends. 
However, their performance remains limited by the absence of fine-grained environmental semantics, which restricts their ability to correct absolute deviations caused by local blockage and heterogeneous structures. 
The vision-based ResNet-18 (C3) captures environmental context through visual semantics and achieves a comparable absolute error, but exhibits a higher relative error, implying that visual information alone lacks reliable geometric scaling in distance-sensitive mmWave propagation. 
By jointly exploiting topology-aware geometric reasoning and visual semantic information, the proposed MM-ResGNN consistently outperforms all uni-modal baselines across different error metrics, demonstrating its capability to accurately model both absolute attenuation levels and relative variations under complex propagation conditions.

\begin{table}[htbp]
\caption{Path Loss Prediction Performance Comparison}
\label{tab:main_comparison}
\centering
\begin{tabular}{llccc}
\hline
\textbf{ID} & \textbf{Model / Configuration} & \textbf{MAE (dB)} & \textbf{NMSE} & \textbf{MAPE (\%)} \\
\hline
C0 & Physical Baseline & 7.9543 & 0.0145 & 7.0264 \\
\hline
\multirow{3}{*}{C1} 
 % & 3GPP UMi (LoS) & 11.5393 & 0.0239 & 9.7771 \\
 & 3GPP UMi (NLoS) & 12.8960 & 0.0233 & 11.8740 \\
 & ABG model & 12.9244 & 0.0233 & 11.9014 \\
 & FSPL model & 11.5091 & 0.0238 & 9.7506 \\
\hline
C2 & Geometry-driven MLP & 6.8109 & 0.0124 & 5.8538 \\
C3 & Vision-based ResNet-18 & 7.0116 & 0.0130 & 6.0698 \\
C4 & Uni-modal Geometric GNN & 6.5817 & 0.0116 & 5.6846 \\
\hline
A0 & \textbf{MM-ResGNN} & \textbf{5.7991} & \textbf{0.0098} & \textbf{5.0498} \\
\hline
\end{tabular}
\end{table}

\subsection{Ablation Studies and Mechanism Analysis}

\begin{figure}[!t]
    \centering
    \subfloat[Comparison with baseline models]{\includegraphics[width=0.48\textwidth]{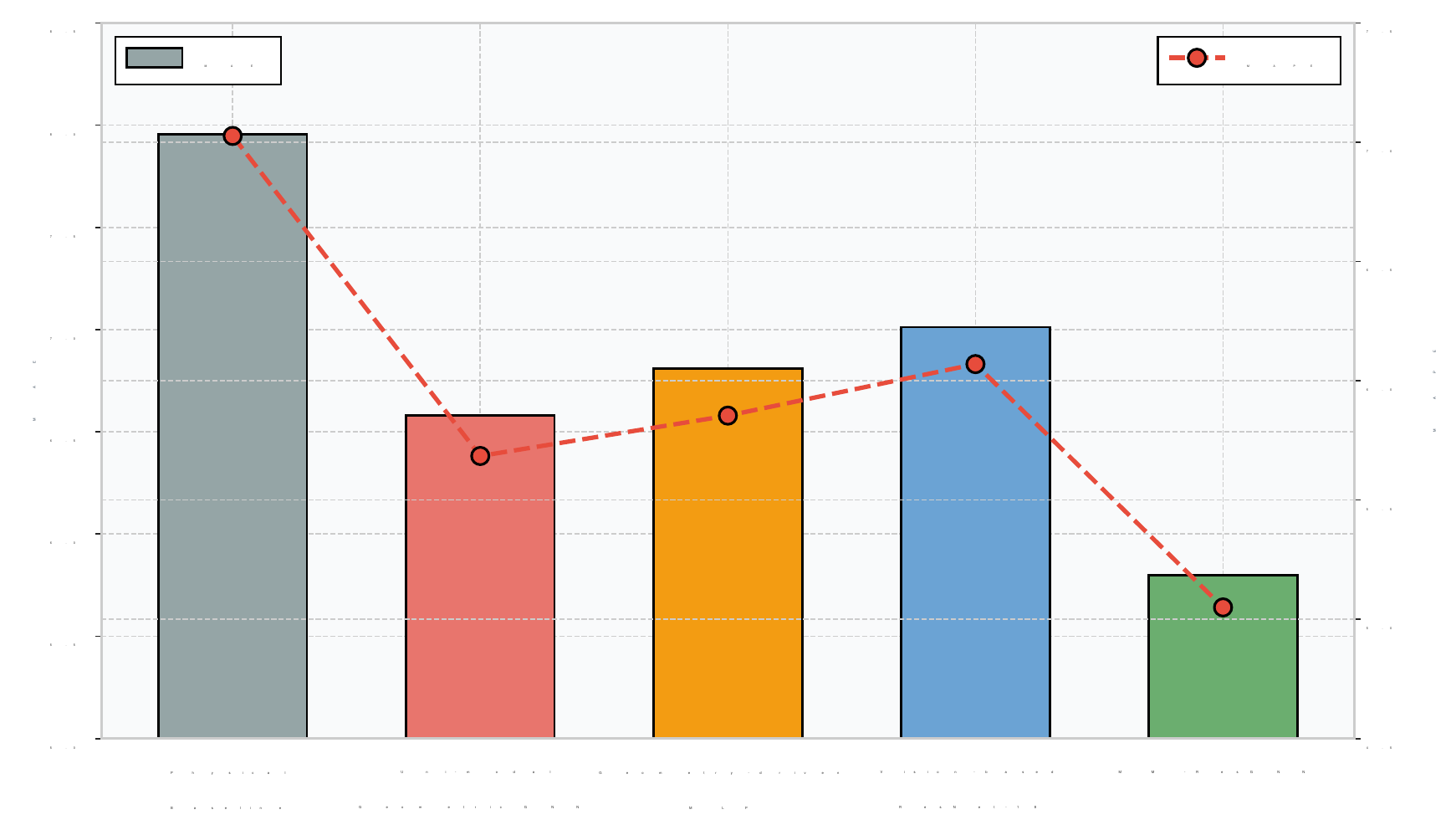}}
    \\
    \subfloat[Ablation study of internal modules]{\includegraphics[width=0.48\textwidth]{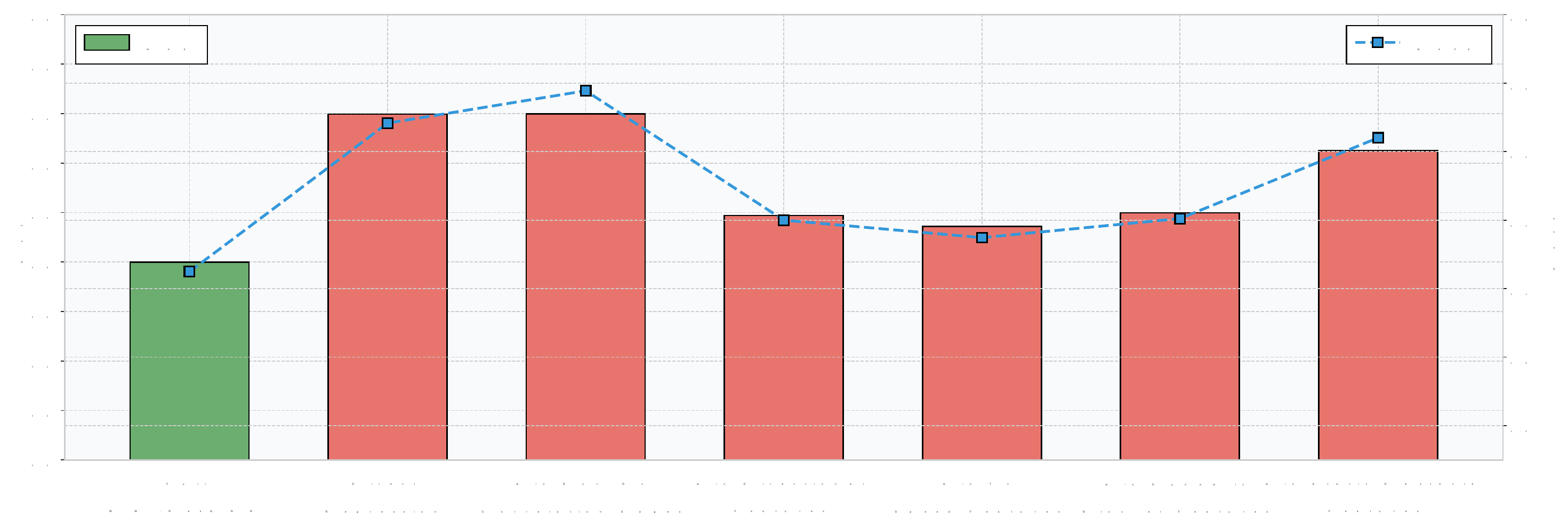}}
    \caption{Quantitative performance analysis in the Urban Wide Lane scenario. (a) Comparison between the physical baseline, single-modal variants, and the proposed MM-ResGNN. (b) Impact of individual components on prediction accuracy, highlighting the significance of residual learning and topological correlations.}
    \label{fig:performance_analysis}
\end{figure}

A systematic ablation study is conducted to quantify the contributions of different architectural components and feature design choices in the proposed MM-ResGNN. The evaluation focuses on three aspects, including the selection of GNN operators, the role of residual learning and collaborative topology, and the impact of physics-aware feature engineering. Quantitative results are summarized in Table~\ref{tab:ablation}.
The corresponding ablation trends are also illustrated in Fig.~\ref{fig:performance_analysis}(b).

\subsubsection{Analysis of GNN Architectures}

The impact of different GNN operators is evaluated by replacing the Transformer-based convolution with graph convolutional network (GCN) and graph attention network (GAT) variants, denoted as V1 and V2, respectively. As reported in Table~\ref{tab:ablation}, the GCN-based variant results in an MAE of 6.47~dB. The isotropic aggregation mechanism of GCN assigns equal importance to neighboring nodes and therefore does not explicitly account for directional propagation characteristics.

% 修改名称适应最新的baseline的名称

The GAT-based variant reduces the MAE to 6.16~dB by introducing attention-based weighting among neighbors. However, edge attributes are not directly incorporated into the attention computation, which limits the ability to exploit physics-aware channel features. In contrast, the Transformer-based operator achieves the lowest MAE by integrating high-dimensional edge features into the attention mechanism, enabling anisotropic aggregation that is consistent with the directional and blockage-sensitive nature of mmWave propagation.

\subsubsection{Impact of Residual Learning and Collaborative Topology}

The effect of residual learning is evaluated by comparing the proposed MM-ResGNN with a direct regression variant without baseline decoupling, denoted as A1. As shown in Table~\ref{tab:ablation}, the removal of residual learning increases the MAE from 5.80~dB to 6.40~dB. This performance degradation indicates that separating large-scale deterministic attenuation from stochastic residual components simplifies the learning objective and improves optimization stability.

The role of collaborative topology is examined by removing Rx--Rx correlation edges, resulting in variant A2. The MAE increases to 6.40~dB under this configuration. The observed degradation suggests that spatial correlations among neighboring Rxs provide complementary contextual information, which contributes to enforcing spatial consistency in path loss prediction across the local region.

\subsubsection{Contribution of Physics-Aware Feature Engineering}

The contribution of individual feature groups is analyzed through variants A3 to A7. Among all feature categories, static obstacle features associated with buildings and trees are observed to be the most influential. As reported in Table~\ref{tab:ablation}, removing static obstacle features increases the MAE to 6.25~dB, indicating that large-scale blockage induced by permanent structures dominates mmWave path loss variation in dense urban environments. Directional features and dynamic blocker features also contribute to prediction accuracy. Excluding directional features in A3 increases the MAE to 5.99~dB, while removing dynamic blocker features in A5 results in an MAE of 6.00~dB. These results suggest that angular alignment and transient occlusions play a non-negligible role in characterizing localized NLoS conditions. The Tx speed feature, evaluated in A4, leads to a moderate performance degradation with an MAE of 5.94~dB. Although the impact is less pronounced than static blockage features, motion-related information provides complementary cues for modeling time-varying shadowing effects.

The effectiveness of the proposed grouped edge encoder is validated by comparing A0 with the flat MLP-based encoder A7. The grouped encoding strategy achieves lower error by processing heterogeneous physical quantities in separate embedding subspaces, which mitigates the dominance of large-scale features, such as distance and preserves the contribution of fine-grained propagation factors.
Overall, the ablation results demonstrate that accurate mmWave path loss prediction benefits from the joint consideration of anisotropic topological aggregation, residual-based physical decoupling, and structured encoding of heterogeneous propagation features.

\begin{table}[htbp]
\caption{Ablation Study and Architectural Analysis Results}
\label{tab:ablation}
\centering
\begin{tabular}{l p{3cm} ccc}
\hline
\textbf{ID} & \textbf{Ablation / Configuration} & \textbf{MAE (dB)} & \textbf{NMSE} & \textbf{MAPE (\%)} \\
\hline
A0 & \textbf{Full MM-ResGNN} & \textbf{5.7991} & \textbf{0.0098} & \textbf{5.0498} \\
\hline
\multicolumn{2}{l}{\textit{GNN Operator Analysis}} \\
V1 & GCN-based multi-modal variant & 6.4657 & 0.0115 & 5.6132 \\
V2 & GAT-based multi-modal variant & 6.1579 & 0.0104 & 5.3465 \\
\hline
\multicolumn{2}{l}{\textit{Mechanism \& Topology Ablation}} \\
A1 & Direct Regression & 6.3980 & 0.0117 & 5.4831 \\
A2 & w/o Rx--Rx Correlation Edges & 6.3981 & 0.0115 & 5.5778 \\
\hline
\multicolumn{2}{l}{\textit{Feature Engineering Ablation}} \\
A3 & w/o Directional features & 5.9882 & 0.0101 & 5.2000 \\
A4 & w/o Tx speed feature & 5.9437 & 0.0101 & 5.1489 \\
A5 & w/o Dynamic blocker features & 5.9987 & 0.0103 & 5.2047 \\
A6 & w/o Static obstacle features & 6.2503 & 0.0105 & 5.4404 \\
A7 & Flat Edge Encoder & 5.9990 & 0.0102 & 5.2182 \\
\hline
\end{tabular}
\end{table}

Overall, the ablation studies provide a comprehensive validation of the proposed MM-ResGNN design. 
The results consistently show that no single component alone is sufficient to achieve accurate mmWave path loss prediction in complex vehicular environments. 
Anisotropic graph aggregation is essential for modeling directional and blockage-sensitive propagation, residual learning effectively stabilizes optimization by decoupling deterministic attenuation from stochastic variations, and physics-aware feature engineering enables the network to exploit heterogeneous propagation cues in a structured manner. 
The observed performance degradation in all ablated variants confirms that these components play complementary roles and must be jointly considered to accurately characterize large-scale path loss under realistic propagation conditions.

\subsection{Fusion Strategy and Backbone Efficiency}
% The integration of multi-modal features is further analyzed in Table~\ref{tab:fusion_eff}. 
The strategic selection of cross-modal integration mechanisms and visual feature extractors is critical for optimizing the balance between prediction fidelity and inference efficiency. Table~\ref{tab:fusion_eff} summarizes a series of comparative experiments designed to validate the efficacy of the proposed gated fusion strategy and the choice of the convolutional backbone.

% 语言修改
\subsubsection{Cross-Modal Fusion Mechanisms}

The influence of different cross-modal fusion strategies is evaluated by comparing gated fusion, feature concatenation, and cross-attention-based fusion. Quantitative results are summarized in Table~\ref{tab:fusion_eff}. Feature concatenation results in an MAE of 6.04~dB, indicating limited effectiveness in jointly modeling heterogeneous geometric and visual representations.
Cross-attention-based fusion achieves an MAE of 5.81~dB, which is comparable to the performance obtained with gated fusion. However, cross-attention introduces additional computational overhead due to pairwise attention operations between modalities. In contrast, gated fusion achieves an MAE of 5.80~dB while maintaining a lightweight structure and enabling adaptive regulation of geometric and visual contributions.
The results indicate that explicitly modeling modality importance through a gating mechanism provides an effective balance between prediction accuracy and computational efficiency.

\subsubsection{Visual Backbone Selection}

The effect of visual backbone depth is examined by replacing the ResNet-18 backbone with a deeper ResNet-50 architecture. As reported in Table~\ref{tab:fusion_eff}, the ResNet-18 backbone achieves an MAE of 5.80~dB, whereas the ResNet-50 backbone yields a slightly higher MAE of 5.88~dB.
The observed performance difference indicates that mmWave path loss prediction primarily benefits from visual cues related to coarse structural layouts, blockage contours, and material boundaries. Such propagation-relevant information is effectively captured by shallower convolutional layers, while deeper semantic abstractions do not provide additional discriminative gains for this task.
These observations confirm that lightweight visual backbones are sufficient for extracting environmental semantics relevant to mmWave propagation, enabling efficient multi-modal learning without unnecessary model complexity.

\begin{table}[htbp]
\caption{Impact of Fusion Strategy and Visual Backbone Selection}
\label{tab:fusion_eff}
\centering
\begin{tabular}{llccc}
\hline
\textbf{ID} & \textbf{Configuration} & \textbf{MAE (dB)} & \textbf{NMSE} & \textbf{MAPE (\%)} \\
\hline
A0 & \textbf{Gated Fusion} & \textbf{5.7991} & \textbf{0.0098} & \textbf{5.0498} \\
A8 & Concatenation Fusion & 6.0413 & 0.0103 & 5.2657 \\
A9 & Cross-Attention Fusion & 5.8123 & 0.0099 & 5.0491 \\
A10 & ResNet-50 Backbone & 5.8794 & 0.0101 & 5.1008 \\
% A11 & EfficientNet-B0 Backbone & 5.8760 & 0.0100 & 5.1166 \\
\hline
\end{tabular}
\end{table}

Overall, the results in Table~\ref{tab:fusion_eff} demonstrate that both the fusion strategy and the choice of visual backbone play a critical role in balancing prediction accuracy and computational efficiency. 
Explicitly modeling modality importance through gated fusion enables effective integration of geometric and visual representations without introducing excessive computational overhead. 
Meanwhile, the comparable or inferior performance of deeper visual backbones indicates that mmWave path loss prediction primarily relies on coarse structural and blockage-related visual cues rather than high-level semantic abstractions. 
These observations confirm that a lightweight gated fusion mechanism combined with a shallow convolutional backbone provides an efficient and effective design choice for multi-modal mmWave path loss modeling.

\subsection{Cross-Scenario Generalization and Few-Shot Adaptation}

Cross-scenario generalization capability is evaluated to assess the robustness and practical applicability of the proposed MM-ResGNN under domain shifts. The model is pre-trained on the urban wide lane scenario and transferred to two target scenarios with distinct topological and environmental characteristics, i.e., the urban crossroad scenario and the suburban forking road scenario. The evaluation focuses on both full-data transfer performance and data-efficient adaptation under limited target-domain supervision.

\subsubsection{Benchmark of Basic Generalization Capability}

\begin{figure}[!t]
    \centering
    \includegraphics[width=0.48\textwidth]{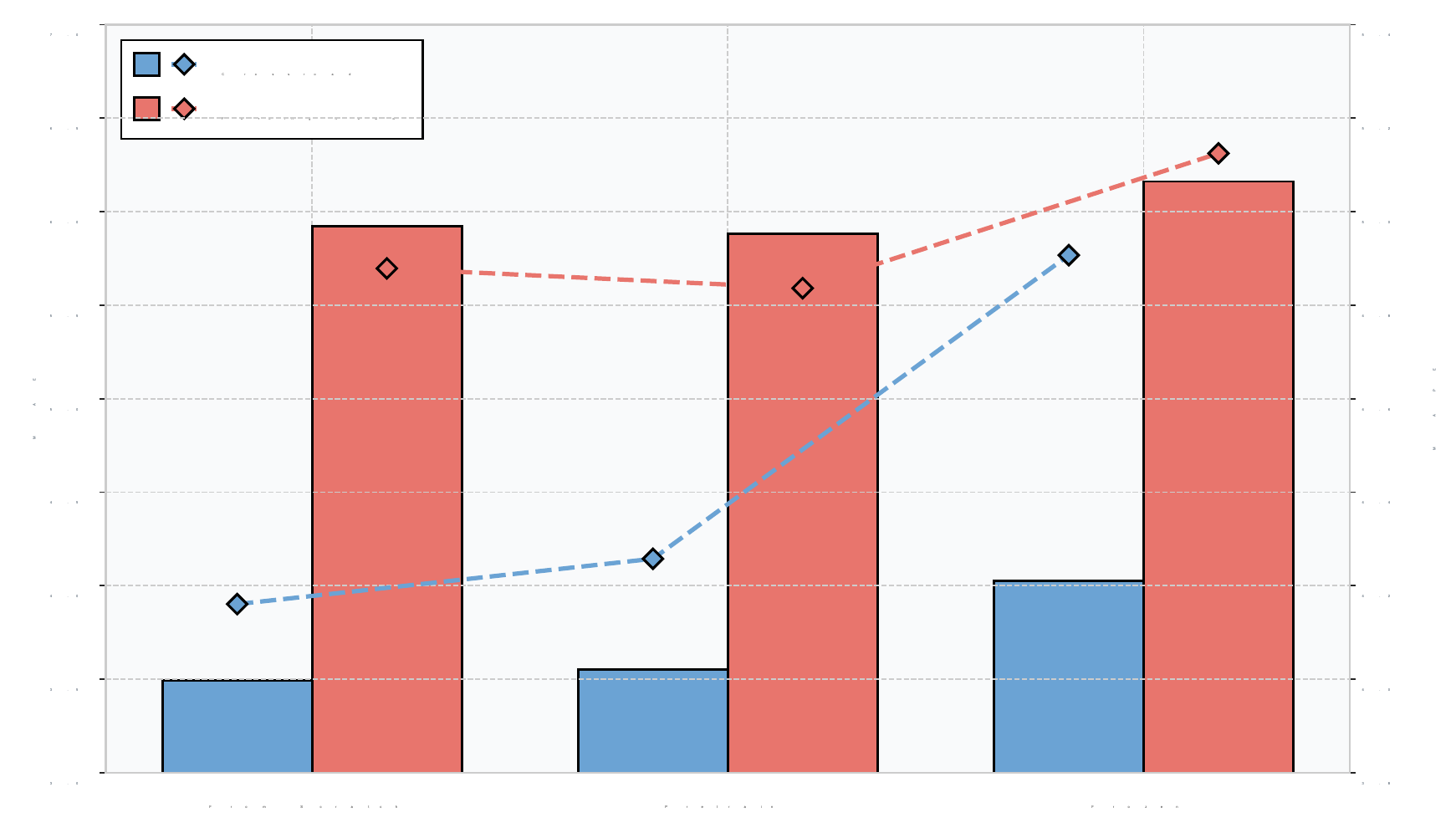} % 这里对应你提到的柱状+折线混合图
    \caption{Benchmark of generalization performance across scenarios. The bar charts represent MAE, while the dashed lines with diamond markers indicate MAPE trends for different transfer strategies.}
    \label{fig:transfer_bar}
\end{figure}

Generalization performance is first evaluated utilizing the complete target-domain datasets. Three transfer strategies are considered under full target-domain supervision to examine different aspects of model generalization and adaptation. 
Training from scratch serves as a reference to evaluate the intrinsic learnability of the target scenario without prior knowledge. 
Full fine-tuning of the pre-trained model utilizing all available target-domain data assesses the transferability of learned representations when sufficient supervision is provided. 
Freezing the pre-trained backbone while retraining only the regression head isolates the contribution of domain-invariant features learned during pre-training. The comparative results are illustrated in Fig.~\ref{fig:transfer_bar} and summarized in Table~\ref{tab:transfer_baseline}.

As reported in Table~\ref{tab:transfer_baseline}, full fine-tuning achieves comparable performance to training from scratch across both target scenarios, and yields a slight gain in the suburban forking road scenario, while being marginally worse in the urban crossroad scenario. In the suburban forking road scenario, the MAE reaches 5.88~dB under full fine-tuning, compared with 5.92~dB obtained by training from scratch. This observation indicates that the representations learned in the urban wide lane scenario retain transferability across different roadway geometries and environmental layouts. When the backbone is frozen, prediction accuracy decreases relative to full fine-tuning. This behavior suggests that a portion of the learned representations captures domain-invariant propagation characteristics, while scenario-specific adaptation primarily benefits from updating higher-level layers.

\subsubsection{Data Efficiency in Few-Shot Adaptation}

\begin{figure}[!t]
    \centering
    \subfloat[Urban crossroad scenario]{\includegraphics[width=0.48\textwidth]{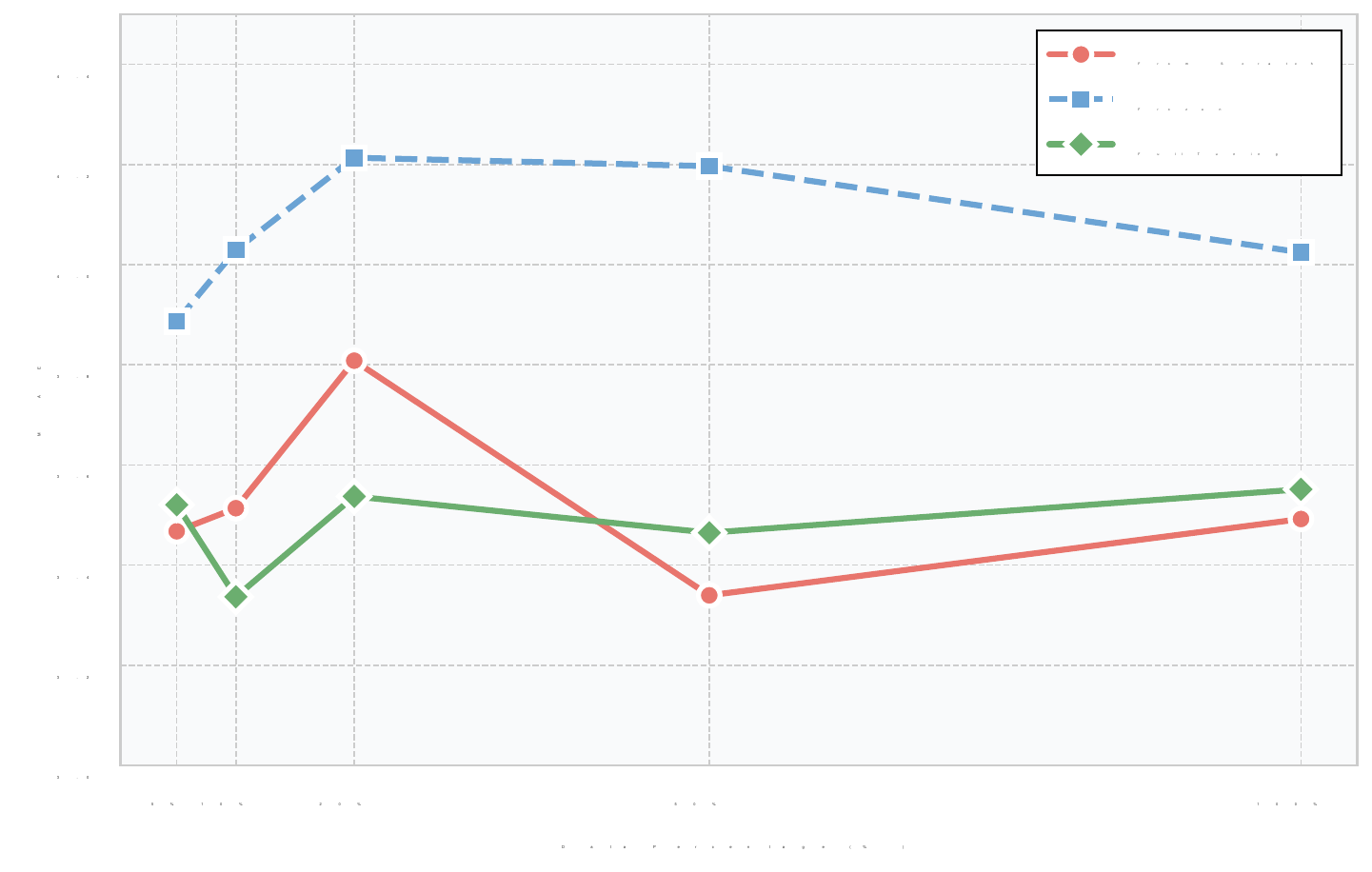}}
    \\
    \subfloat[Suburban forking road scenario]{\includegraphics[width=0.48\textwidth]{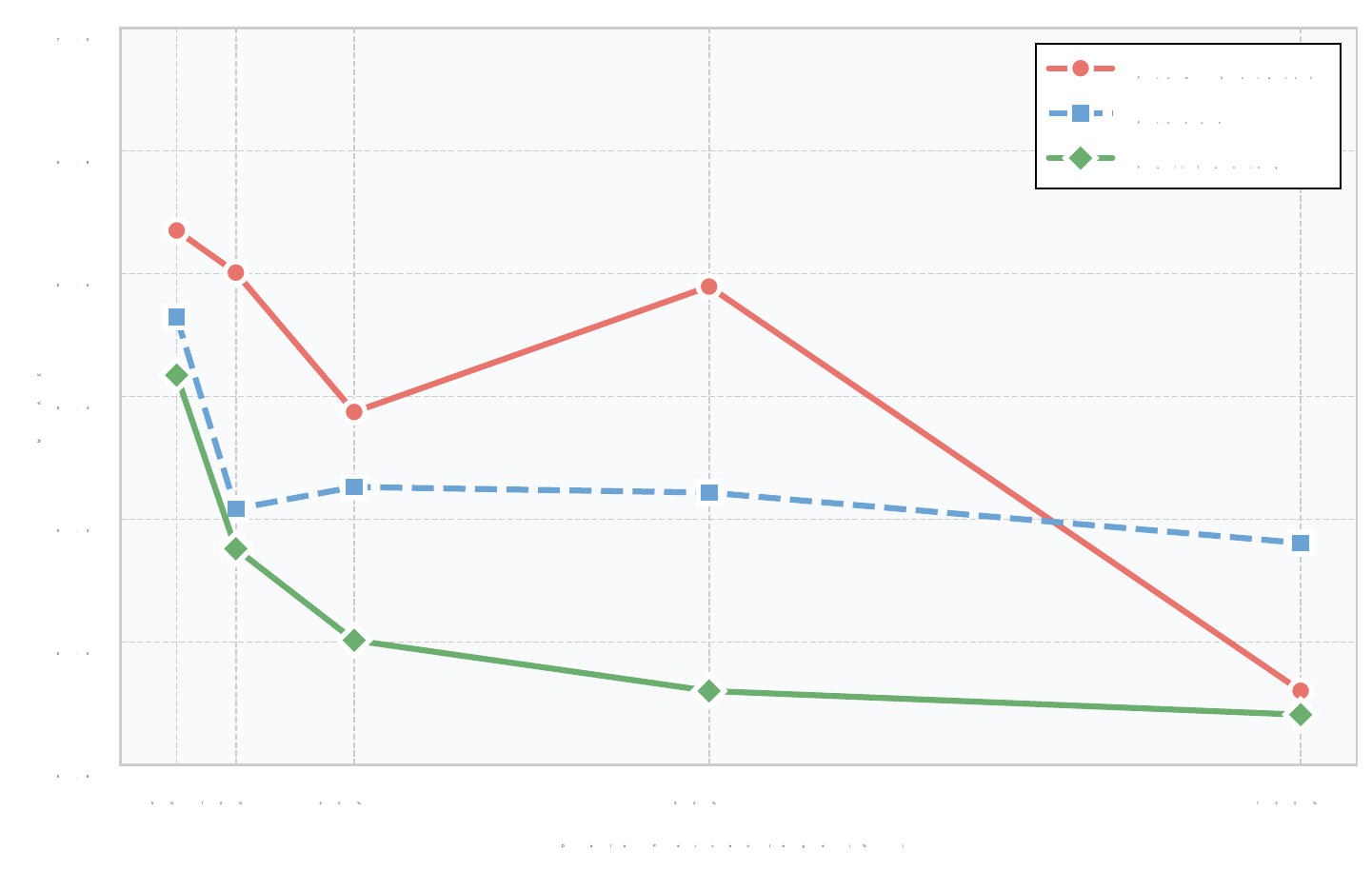}}
    \caption{Data efficiency analysis. The curves illustrate the MAE performance as a function of target data percentage for four training strategies.}
    \label{fig:data_efficiency_curves}
\end{figure}

In contrast to the full-data setting above, few-shot adaptation is evaluated by progressively reducing the proportion of labeled target-domain data from 5\% to 100\%.
 The MAE trends for the urban crossroad and suburban forking road scenarios are illustrated in Fig.~\ref{fig:data_efficiency_curves}, and detailed numerical results are reported in Table~\ref{tab:data_efficiency}.

% --- Table III: 基础泛化能力对比 (100%数据) ---
\begin{table}[htbp]
\caption{Benchmark of Transfer Learning Performance on Target Scenarios (100\% Data)}
\label{tab:transfer_baseline}
\centering
\begin{tabular}{llccc}
\hline
\textbf{Target Scenario} & \textbf{Strategy} & \textbf{MAE (dB)} & \textbf{NMSE} & \textbf{MAPE (\%)} \\
\hline
\multirow{3}{*}{Urban Crossroad} & Scratch & \textbf{3.4920} & \textbf{0.0070} & \textbf{4.1606} \\
 & Pretrain & 3.5514 & 0.0071 & 4.2576 \\
 & Frozen & 4.0245 & 0.0071 & 4.9069 \\
\hline
\multirow{3}{*}{Suburban Forking} & Scratch & 5.9226 & 0.0111 & 4.8787 \\
 & Pretrain & \textbf{5.8817} & \textbf{0.0110} & \textbf{4.8364} \\
 & Frozen & 6.1614 & 0.0114 & 5.1247 \\
\hline
\end{tabular}
\end{table}

\begin{table}[htbp]
\caption{Data Efficiency and Few-Shot Adaptation Performance}
\label{tab:data_efficiency}
\centering
\begin{tabularx}{\columnwidth}{lcXXXX}
\hline
\textbf{Target Scenario} & \textbf{Ratio} & \textbf{Scratch} & \textbf{Frozen} & \textbf{Full Tuning} \\
\hline
\multirow{5}{*}{Urban Crossroad} 
& 5\%  & \textbf{3.4675} & 3.8868  & 3.5225 \\
& 10\% & 3.5136 & 4.0294  & \textbf{3.3366} \\
& 20\% & 3.8082 & 4.2135  & \textbf{3.5371} \\
& 50\% & \textbf{3.3396} & 4.1967  & 3.4646 \\
& 100\% & \textbf{3.4920} & 4.0245  & 3.5514 \\
\hline
\multirow{5}{*}{Suburban Forking} 
& 5\%  & 6.6703 & 6.5289   & \textbf{6.4348} \\
& 10\% & 6.6018 & 6.2170 & \textbf{6.1522} \\
& 20\% & 6.3747 & 6.2525  & \textbf{6.0028} \\
& 50\% & 6.5790 & 6.2434  & \textbf{5.9203} \\
& 100\% & 5.9206 & 6.1614  & \textbf{5.8817} \\
\hline
\end{tabularx}
\end{table}

% --- Table VI: 跨场景迁移能力评估 (10%数据, Frozen模式) ---
\begin{table}[htbp]
\caption{Cross-Domain Transferability Evaluation (10\% Data, Frozen Strategy)}
\label{tab:cross_domain}
\centering
\begin{tabular}{llccc}
\hline
\textbf{Source} & \textbf{Target} & \textbf{MAE (dB)} & \textbf{NMSE} & \textbf{MAPE (\%)} \\
\hline
Wide Lane & Crossroad & \textbf{3.9917} & \textbf{0.0072} & \textbf{4.8445} \\
Wide Lane & Forking & 6.2187 & 0.0111 & 5.1973 \\
Crossroad & Forking & 6.7614 & 0.0110 & 5.7830 \\
Forking & Crossroad & 4.0344 & 0.0071 & 4.9048 \\
\hline
\end{tabular}
\end{table}

The performance comparison in Table~\ref{tab:data_efficiency} reveals distinct adaptation behaviors across the two target scenarios. In the suburban forking road scenario, the full-tuning strategy consistently outperforms the scratch-based approach across all data ratios. Notably, at the 5\% data threshold, full-tuning achieves an MAE of 6.43~dB, while the scratch-based model yields a significantly higher error of 6.67~dB. Even the frozen backbone strategy provides a more accurate prediction, i.e., 6.53~dB, than training from scratch in this low-data regime. Such a trend indicates that the irregular scattering and sparse vegetation characteristics of the suburban environment are inherently difficult to characterize with limited samples. The representations learned from the propagation-rich wide lane scenario serve as a robust physical prior, effectively regularizing the model and preventing over-fitting under sparse supervision.

In the urban crossroad scenario, the scratch-based model exhibits surprisingly competitive performance, particularly at the 5\% and 50\% data ratios. This behavior suggests that the highly structured Manhattan-grid geometry and the resulting street-canyon effects are relatively easy for the GNN to internalize even with a minimal number of samples. However, the full-tuning strategy still achieves the optimal MAE of 3.34~dB at the 10\% ratio, surpassing the scratch-based model's 3.51~dB. The marginal performance gain in the crossroad scenario, compared to the suburban scenario, implies that the ``corner effects" and intersection-specific shadowing patterns require highly localized optimization, reducing the relative importance of source-domain knowledge.

Overall, the data efficiency analysis confirms that pre-training provides a more stable and accurate initialization for optimization, especially in environments with high propagation complexity. While structured urban scenes allow for rapid learning from scratch, the transfer of MM-ResGNN is essential for maintaining high-fidelity path loss prediction in irregular and data-constrained scenarios.

\subsubsection{Analysis of Cross-Domain Transferability}

Cross-domain transferability among different source–target pairs is quantified in Table~\ref{tab:cross_domain}. Transferring from the urban wide lane scenario to the urban crossroad scenario achieves an MAE of 3.99~dB, which is substantially lower than the MAE of 6.22~dB observed when transferring to the suburban forking road scenario. The observed performance difference is consistent with the underlying propagation environments. Urban scenarios share similar building materials, street-canyon structures, and multipath characteristics, which facilitates feature reuse across domains. In contrast, the suburban forking road scenario is dominated by vegetation scattering and ground reflections, resulting in a propagation environment that differs significantly from dense urban settings.
These results indicate that source-domain selection plays a critical role in transfer learning for machine synesthesia-based channel modeling, and that pre-training on propagation-rich urban environments provides more transferable representations.

Overall, the cross-scenario experiments demonstrate that the proposed MM-ResGNN learns transferable representations that generalize effectively across heterogeneous vehicular environments. 
Under full target-domain supervision, pre-training provides performance comparable to or better than training from scratch, indicating that the learned representations capture domain-invariant propagation characteristics. 
More importantly, under limited-data regimes, MM-ResGNN exhibits clear advantages in data efficiency, particularly in complex suburban environments where propagation mechanisms are difficult to learn from sparse samples. 
The cross-domain transfer results further suggest that pre-training on propagation-rich urban scenarios yields more reusable representations, highlighting the importance of source-domain selection for practical deployment. 
Together, these findings confirm that MM-ResGNN offers robust generalization and adaptation capabilities for mmWave path loss prediction under realistic domain shifts.

\section{Conclusions}
This paper studies mmWave path loss prediction for vehicular scenarios utilizing the VMMPL dataset, which achieves precise alignment between RGB images and global semantic information in the physical space, and link-level RT-based path loss data in the electromagnetic space across three representative vehicular scenarios. Based on an ESPL-Graph representation, MM-ResGNN combines a geometry-driven physical baseline with residual learning, and integrates topology-aware graph features with visual semantics via gated fusion. Simulation results show that MM-ResGNN outperforms empirical models and uni-modal baselines. In the urban wide lane scenario, MM-ResGNN achieves an MAE of 5.7991~dB and an NMSE of 0.0098. Cross-scenario experiments further indicate that the learned representations can be adapted to new environments utilizing few-shot fine-tuning, improving data efficiency when target-domain labels are limited. Future work will investigate deployment-oriented extensions, such as beam management and environment-aware network optimization, as well as the impact of varying Rx density and graph size under practical sensing constraints.

% \section{References Section}
% You can use a bibliography generated by BibTeX as a .bbl file.
%  BibTeX documentation can be easily obtained at:
%  http://mirror.ctan.org/biblio/bibtex/contrib/doc/
%  The IEEEtran BibTeX style support page is:
%  http://www.michaelshell.org/tex/ieeetran/bibtex/
 
 % argument is your BibTeX string definitions and bibliography database(s)
%\bibliography{IEEEabrv,../bib/paper}

\begin{thebibliography}{1}
\bibliographystyle{IEEEtran}
% \bibitem{6ggg}
% A. Molisch, \emph{Wireless Communications}. UK: John Wiley Sons, 2011.
% 6G V2X

\bibitem{ITS1}
F. Zhu, Z. Li, S. Chen, and G. Xiong, ``Parallel transportation management and control system and its applications in building smart cities,'' \emph{IEEE Trans. Intell. Transp. Syst.}, vol. 17, no. 6, pp. 1576--1585, Jun. 2016.

\bibitem{6G_V2X_Review}
M. Noor-A-Rahim \textit{et al.}, ``6G for vehicle-to-everything (V2X) communications: Enabling technologies, challenges, and opportunities,'' \emph{Proc. IEEE}, vol. 110, no. 6, pp. 712--734, Jun. 2022.

\bibitem{6G_V2X_Slicing}
A. Boualouache, A. A. Jolfaei, and T. Engel, ``Multi-process federated learning with stacking for securing 6G-V2X network slicing at cross-borders,'' \emph{IEEE Trans. Intell. Transp. Syst.}, vol. 25, no. 9, pp. 10941--10952, Sept. 2024.

\bibitem{6G_Enabled_ATS}
R. Liu \textit{et al.}, ``6G enabled advanced transportation systems,'' \emph{IEEE Trans. Intell. Transp. Syst.}, vol. 25, no. 9, pp. 10564--10580, Sep. 2024.

\bibitem{channel_1}
Y. Niu \textit{et al.}, ``Interference management
for integrated sensing and communication systems: A survey,'' \emph{IEEE
Internet Things J.}, vol. 12, no. 7, pp. 8110--8134, Apr. 2025.


\bibitem{PL_application1}
R. Borralho, A. Mohamed, A. U. Quddus, P. Vieira, and R. Tafazolli, ``A
survey on coverage enhancement in cellular networks: Challenges and
solutions for future deployments," \emph{IEEE Commun. Surveys Tuts.}, vol. 23, no. 2, pp. 1302--1341, secondquarter 2021.

\bibitem{PL_application2}
R. A. Hussain and W. K. Saad, ``A comprehensive survey of path loss types in different wireless communication environments,'' in \emph{Proc. IEEE I2CACIS'24}, Shah Alam, Malaysia, Aug. 2024, pp. 249--255.

\bibitem{PL_application3}
B. Zhu, E. Bedeer, H. H. Nguyen, R. Barton, and Z. Gao, ``UAV trajectory
planning for AoI-minimal data collection in UAV-aided IoT networks by
Transformer," \emph{IEEE Trans. Wireless Commun.}, vol. 22, no. 2, pp. 1343--1358, Feb. 2023.


\bibitem{PL_tx_rx}
C. Hazirbas, L. Ma, C. Domokos, and D. Cremers, ``FuseNet: Incorporating depth into semantic segmentation via fusion-based CNN architecture,'' in \emph{Proc. ACCV'16}, Taipei, Taiwan, Mar. 2016, pp. 213--228.

\bibitem{CI_model}
T. S. Rappaport \textit{et al.}, ``Overview of millimeter wave communications for fifth-generation (5G) wireless networks-with a focus on propagation models," \emph{IEEE Trans. Antennas Propag.}, vol. 65, no. 12, pp. 6213-–6230, Dec. 2017.


\bibitem{empirical_model2}
X. Cai et al., ``An empirical air-to-ground channel model based on passive measurements in LTE," \emph{IEEE Trans. Veh. Technol.}, vol. 68, no. 2, pp. 1140--1154, Feb. 2019.

\bibitem{empirical_model3}
W. Tang et al., ``Path loss modeling and measurements for reconfigurable
intelligent surfaces in the millimeter-wave frequency band," \emph{IEEE Trans. Commun}, vol. 70, no. 9, pp. 6259--6276, Sept. 2022.

\bibitem{PathLoss_Urban_ML}
I. F. M. Rafie, S. Y. Lim, and M. J. H. Chung, ``Path loss prediction in urban areas: A machine learning approach,'' \emph{IEEE Antennas Wireless Propag. Lett.}, vol. 22, no. 4, pp. 809--813, Apr. 2023.
\bibitem{PathLoss_Manhattan}
A. Gupta \textit{et al.}, ``Machine learning-based urban canyon path loss prediction using 28 GHz Manhattan measurements,'' \emph{IEEE Trans. Antennas Propag.}, vol. 70, no. 6, pp. 4096--4111, Jun. 2022.
\bibitem{PathLoss_RT_Sionna}
G. Xia \textit{et al.}, ``Path loss prediction in urban environments with Sionna-RT based on accurate propagation scene models at 2.8 GHz,'' \emph{IEEE Trans. Antennas Propag.}, vol. 72, no. 10, pp. 7986--7997, Oct. 2024.
% Conventional ML
\bibitem{SVM_RF_INDOOR}
Y. Nuñez, L. Lovisolo, L. da Silva Mello, and C. Orihuela, ``On the interpretability of machine learning regression for path-loss prediction of millimeter-wave links,'' \emph{Expert Syst. Appl.}, vol. 215, no. 14, p. 119324, Nov. 2022.
\bibitem{V2V_SVM_RF}
Y. Nuñez \textit{et al.}, ``Path loss prediction for vehicular-to-infrastructure communication using machine learning techniques," in \emph{Proc. IEEE VCC'23}, New York, America, Nov. 2023, pp. 270--275.

\bibitem{CNN_Satellite}
O. Ahmadien, H. F. Ates, T. Baykas, and B. K. Gunturk, ``Predicting path loss distribution of an area from satellite images using deep learning,'' \emph{IEEE Access}, vol. 8, no. 1, pp. 64982--64991, Apr. 2020.
\bibitem{CNN_City_Map}
K. Inoue, K. Ichige, T. Nagao, and T. Hayashi, ``Learning-based prediction method for radio wave propagation using images of building maps,'' \emph{IEEE Antennas Wireless Propag. Lett.}, vol. 21, no. 1, pp. 124--128, Oct. 2022.
\bibitem{SatImage_PL_Prediction}
C. Wang \textit{et al.}, ``Channel path loss prediction using satellite images: A deep learning approach,'' \emph{IEEE Trans. Mach. Learn. Commun. Netw.}, vol. 2, no. 1, pp. 1357--1368, Dec. 2024.

% GNN General
\bibitem{GNN_IoT}
T. Chen \textit{et al.}, ``A GNN-based supervised learning framework for resource allocation in wireless IoT networks,'' \emph{IEEE Internet Things J.}, vol. 9, no. 3, pp. 1712--1724, Jun. 2021.
\bibitem{GNN_EdgeUpdate}
Y. Wang, Y. Li, Q. Shi, \textit{et al.}, ``ENGNN: A general edge-update empowered GNN architecture for radio resource management in wireless networks,'' \emph{IEEE Trans. Wireless Commun.}, vol. 23, no. 6, pp. 5330--5344, Oct. 2023.
\bibitem{GNN_Routing}
Y. Huang \textit{et al.}, ``A GNN-enabled multipath routing algorithm for spatial-temporal varying LEO satellite networks,'' \emph{IEEE Trans. Veh. Technol.}, vol. 73, no. 4, pp. 5454--5468, Apr. 2024.
\bibitem{GNN_OAC}
Y. Yang, Z. Zhang, Y. Tian, R. Jin, and C. Huang, ``Implementing graph neural networks over wireless networks via over-the-air computing: A joint communication and computation framework,'' \emph{IEEE Wireless Commun.}, vol. 30, no. 3, pp. 62--69, Jun. 2023.
% GNN Path Loss & Channel
\bibitem{GNN_PathLoss_Geo}
X. Liu, K. Guo, R. Sun, \textit{et al.}, ``A path loss prediction scheme based on graph neural networks and geographical information,'' in \emph{Proc. ETAI'25}, Harbin, China, Jul. 2025, pp. 1--6.
\bibitem{GNN_ChannelMod}
X. Wang \textit{et al.}, ``Graph neural network enabled propagation graph method for channel modeling,'' \emph{IEEE Trans. Veh. Technol.}, vol. 73, no. 9, pp. 12280--12289, Sept. 2024.

\bibitem{SoM}
X. Cheng \emph{et al.}, ``Intelligent multi-modal sensing-communication integration: Synesthesia of Machines,'' \emph{IEEE Commun. Surveys Tuts.}, vol. 26, no. 1, pp. 258--301, firstquarter 2024.

\bibitem{MMICM}
L. Bai, Z. Huang, M. Sun, X. Cheng, and L. Cui, ``Multi-modal intelligent channel modeling: A new modeling paradigm via Synesthesia of
Machines," \emph{IEEE Commun. Surveys Tuts.}, vol. 28, pp. 2612--2649, Apr. 2025.

\bibitem{WCM}
L. Bai, Z. Han, X. Cai and X. Cheng, ``Multi-modal intelligent channel modeling framework for 6G-enabled networked intelligent systems," \textit{IEEE Wireless Commun.}, early access, 2026, doi: 10.1109/MWC.2025.3624714. 

\bibitem{SynthSoM}
X. Cheng et al., ``SynthSoM: A synthetic intelligent multi-modal
sensing-communication dataset for Synesthesia of Machines (SoM)",  \emph{Sci. Data}, vol. 12, pp. 819–-833, May 2025.

\bibitem{AirSim}
S. Shah, D. Dey, C. Lovett, and A. Kapoor, ``AirSim: High-fidelity visual and physical simulation for autonomous vehicles,'' in \emph{Field and Service Robotics}, M. Hutter and R. Siegwart, Eds. Cham, Switzerland: Springer, Nov. 2017, pp. 621--635.
\bibitem{WI}
\emph{Remcom.} Wireless InSite. [Online]. Available: https://www.remcom.com/wireless-insite-em-propagation-software [Publication date: Jan. 2017, Accessed date: Mar. 2022].

\bibitem{KNN}
P. Cunningham and S. J. Delany, ``k-cearest ceighbour classifiers: A tutorial,'' \emph{ACM Comput. Surv.}, vol. 54, no. 6, pp. 1--25, Jul. 2022.

\bibitem{GTN_NeurIPS19}
S. Yun, M. Jeong, R. Kim, J. Kang, and H. J. Kim, ``Graph transformer networks,'' 
in \emph{Proc. NeurIPS'19}, Vancouver, Canada, Dec. 2019, pp. 11983--11993.

\bibitem{Resnet}
W. Xu, Y.-L. Fu, and D. Zhu, ``ResNet and its application to medical image processing: Research progress and challenges,'' 
\emph{Comput. Methods Programs Biomed.}, vol. 240, Art. no. 107660, Jun. 2022.

\bibitem{CI}
 T. S. Rappaport \emph{et al.}, ``Overview of millimeter wave communications for fifth-generation (5G) wireless networks-with a focus on propagation models," \emph{IEEE Trans. Antennas Propag.}, vol. 65, no. 12, pp. 6213--6230, Dec. 2017.

\bibitem{3GPP}
\emph{Technical Specification Group Radio Access Network; Study on Channel Model for Frequencies From 0.5 to 100 GHz (Release 14)}, document TR 38.901 Version 14.2.0, 3GPP, Sep. 2017. [Online]. Available: http://www.3gpp.org/DynaReport/38901.htm

\bibitem{Rappaport2015ABG}
T. S. Rappaport, G. R. MacCartney, M. K. Samimi, and S. Sun, ``Wideband millimeter-wave propagation measurements and channel models for future wireless communication system design,'' 
\emph{IEEE Trans. Commun.}, vol. 63, no. 9, pp. 3029--3056, Sept. 2015.


\end{thebibliography}
%
\section{Simple References}
You can manually copy in the resultant .bbl file and set second argument of $\backslash${\tt{begin}} to the number of references
 (used to reserve space for the reference number labels box).

\newpage

\vfill

\end{document}